\documentclass[
aps, prd, reprint,
10pt, notitlepage, a4paper,
floats, floatfix,
amsmath, amssymb, amsfonts,
superscriptaddress,
showpacs, showkeys,
nofootinbib,
longbibliography,
]{revtex4-1}

\usepackage{graphicx} 
\usepackage{grffile}
\usepackage[usenames,dvipsnames]{xcolor}
\usepackage{xspace} 
\usepackage{bm} 
\usepackage{amsmath,amsfonts,amssymb,amsthm}
\usepackage[utf8]{inputenc} 
\usepackage{tabularx}
\usepackage[normalem]{ulem}
\usepackage{comment}
\usepackage{multirow}

\xdefinecolor{mylinkcolor}{rgb}{0,0,0.5}
\usepackage[
	bookmarksnumbered, bookmarksopen, bookmarksopenlevel=2,
	breaklinks=true,
	colorlinks=true, filecolor=mylinkcolor, citecolor=mylinkcolor,
	linkcolor=mylinkcolor, urlcolor=mylinkcolor, menucolor=mylinkcolor,
]{hyperref}

\allowdisplaybreaks

\usepackage[dvipsnames]{xcolor}

\newcommand{\Ord}{\mathcal{O}}

\newcommand{\io}{\iota}

\newcommand{\FTI}{{\texttt{FTI}}}
\newcommand{\SEOBNRHM}{{\texttt{SEOBNRHM}}}
\newcommand{\SEOBNR}{{\texttt{SEOBNR}}}
\newcommand{\SEOBNRT}{{\texttt{SEOBNRT}}}
\newcommand{\pSEOBNRHM}{{\texttt{pSEOBNRHM}}}
\newcommand{\pSEOBNR}{{\texttt{pSEOBNR}}}
\newcommand{\pSEOBNRT}{{\texttt{pSEOBNRT}}}
\newcommand{\PhenomX}{{\texttt{PhenomX}}}
\newcommand{\pPhenomX}{{\texttt{pPhenomX}}}
\newcommand{\TIGER}{{\texttt{TIGER}}}

\newcommand{\AEI}{\affiliation{Max Planck Institute for Gravitational Physics (Albert Einstein Institute), Am M\"uhlenberg 1, Potsdam 14476, Germany}}
\newcommand{\Maryland}{\affiliation{Department of Physics, University of Maryland, College Park, MD 20742, USA}}

\begin{document}

\title{Tests of General Relativity with Gravitational-Wave Observations \\ using a Flexible--Theory-Independent Method}

\author{Ajit Kumar Mehta}\email{ajit.mehta@aei.mpg.de}\AEI
\author{Alessandra Buonanno}\email{alessandra.buonanno@aei.mpg.de}\AEI\Maryland
\author{Roberto Cotesta}\email{rcotest1@jhu.edu} \AEI 
\author{Abhirup Ghosh}\email{abhirup.ghosh@aei.mpg.de}\AEI
\author{Noah Sennett}\AEI\Maryland
\author{Jan Steinhoff}\email{jan.steinhoff@aei.mpg.de}\AEI

\date{\today}

\begin{abstract}
	
   We perform tests of General Relativity (GR) with gravitational waves (GWs) from the inspiral stage of compact binaries using a theory-independent framework, which adds generic phase corrections to each multipole of a GR waveform model in frequency domain.  This method has been demonstrated on LIGO-Virgo observations to provide stringent constraints on post-Newtonian predictions of the inspiral and to assess systematic biases that may arise in such parameterized tests. Here, we detail the anatomy of our framework for aligned-spin waveform models. We explore the effects of higher modes in the underlying signal on tests of GR through analyses of two unequal-mass, simulated binary signals similar to GW190412 and GW190814. We show that the inclusion of higher modes improves both the precision and the accuracy of the measurement of the deviation parameters. Our testing framework also allows us to vary the underlying baseline GR waveform model and the frequency at which the non-GR inspiral corrections are tapered off. We find that to optimize the GR test of high-mass binaries, comprehensive studies would need to be done to determine the best choice of the tapering frequency as a function of the binary's properties. We also carry out an analysis on the binary neutron-star event GW170817 to set bounds on the coupling constant $\alpha_0$ of Jordan-Fierz-Brans-Dicke gravity. We take two plausible approaches; the first approach involves translating directly the \emph{theory-agnostic} bound on dipole-radiation into a bound on $\alpha_0$ for different neutron-star equations of state (EOS). The second \emph{theory-specific} approach involves reparameterizing the test such that the deviation parameter is $\alpha_0$ itself. The two approaches provide slightly different bounds, namely, $\alpha_0 \lesssim 2\times 10^{-1}$ and $\alpha_0 \lesssim 4\times 10^{-1}$, respectively, at $68\%$ credible level. These differences arise mainly since in the theory-specific approach the tidal and scalar-charge parameters are fixed coherently for each neutron-star EOS and mass.
  
\end{abstract}

\maketitle

\section{Introduction}\label{FTA:sec:Intro}

Over the past half-decade, observations of gravitational waves (GWs)
have gone from being elusive to the routine. Since the first
detection of GWs in September 2015~\cite{Abbott:2016blz}, the
LIGO~\cite{LIGOScientific:2014pky} and Virgo~\cite{VIRGO:2014yos}
detectors have observed almost a hundred GW
signals~\cite{LIGOScientific:2021djp} from mergers of black holes
(BHs), neutron stars
(NSs)~\cite{TheLIGOScientific:2017qsa,LIGOScientific:2020aai} and
their mixture~\cite{LIGOScientific:2021qlt}. Placed alongside
independent confirmations of these detections, as well as, claims of new
ones~\cite{Nitz:2018imz,Nitz:2019hdf,Venumadhav:2019lyq,Zackay:2019btq,Nitz:2021zwj,Olsen:2022pin},
these results have firmly established the field of GW astronomy.

Besides attempting answers in astrophysics~\cite{LIGOScientific:2017ync, LIGOScientific:2018cki,
  LIGOScientific:2021psn} and cosmology~\cite{LIGOScientific:2021aug,
  LIGOScientific:2021odm} in a manner
complementary to electromagnetic astronomy, GWs are unique probes of
fundamental physics. For more than a century, Albert Einstein's theory
of General Relativity (GR) has been our description of
gravitational interactions, having passed every experimental and
observational challenge, so far. However, for the first time, the
LIGO-Virgo GW observations have allowed us to probe GR in the
large-velocity, highly dynamical, strong-field regime of gravity,
a regime which is inaccessible with tests in the Solar
System~\cite{Will:2014kxa}, in binary pulsars~\cite{Wex:2014nva}, and
with observations around supermassive BHs at the center of
galaxies~\cite{GRAVITY:2018ofz,Do:2019txf,EventHorizonTelescope:2019ths}.

Tests of GR with GW observations come in two distinct flavors:
\emph{theory agnostic} and \emph{theory specific}. The first class of
tests assumes that the underlying GW signal is well-described by GR,
and any potential deviation is characterized by \emph{extra,
  phenomenological}, non-GR degrees of freedom or
\emph{parameters}. These tests use observations of GWs to constrain
the non-GR parameters and check for consistency with their nominal
predictions in GR. While theory-agnostic tests can only comment on
(dis)agreement with GR predictions, the above measurements of
phenomenological non-GR parameters can be translated
to specific modified theories of gravity, albeit there are subtleties, 
as we shall discuss below. Investigations that compare directly the data 
with modified theories of gravity belong to the
  \emph{theory-specific} flavor of tests of GR.

Several tests of GR have been demonstrated using the observations of GW
  signals by the LIGO-Virgo Collaboration (LVC)~\cite{TheLIGOScientific:2016src,TheLIGOScientific:2016pea,Abbott:2018lct,LIGOScientific:2019fpa,LIGOScientific:2020tif,
    LIGOScientific:2021sio}. Among them are the theory-agnostic parameterized tests of the
  \emph{inspiral}, which check for the agreement of the early
  evolutionary (or inspiral) phase of a compact binary coalescence
  composed of BHs and/or NSs with the analytic post-Newtonian (PN)
  approximation for binaries in GR~\cite{Arun:2006hn,Yunes:2009ke,Mishra:2010tp,Li:2011cg,Agathos:2013upa}. Parameterized
  GR waveforms have used the LIGO-Virgo observations to provide
  state-of-the-art bounds on possible deviations from the PN
  predictions~\cite{LIGOScientific:2020tif}. At the same time, these
  theory-agnostic bounds have been used to conduct theory-specific
  tests and constrain particular modified theories of gravity (see,
  e.g., Refs.~\cite{Yunes:2016jcc,Nair:2019iur,Sennett:2019bpc,Nair:2020ggs,Perkins:2021mhb}).

  And yet, parameterized inspiral tests are not all exactly the same;
  they can differ in the underlying GR waveform model, in how the
  non-GR parameters are introduced into it, and in the transition
  beyond the inspiral part. In this work, we develop a framework to
examine how the details of the construction of parameterized waveform
models systematically affect the tests of GR in which they are
employed~\footnote{Parts of this manuscript are based on the PhD
  thesis of Noah Sennett~\cite{Sennett:2021iqr}, in particular Sec.~\ref{FTA:sec:BransDicke} 
follows Chapter 9 of Ref.~\cite{Sennett:2021iqr}.}. {It is also important 
to be able to distinguish deviations from GR due to systematic uncertainties 
of the waveform model from true violations of the theory. This infrastructure
allows us to add generic corrections to the inspiral portion of any 
gravitational waveform, thereby allowing tests of GR
with a broader range of waveform models than previously
possible} (e.g., with the \texttt{TIGER} infrastructure~\cite{Li:2011cg,Agathos:2013upa}).
We call this framework the flexible theory-independent (\FTI{}) approach, and
explore it using synthetic binary BH (BBH) GW signals. In addition,
the conveniently adaptable design of the \FTI{} framework allows us for
easy construction of waveform models for theory-specific tests. As an
example, we apply the \FTI{} framework to the first binary NS
(BNS) merger observed by the LIGO and Virgo detectors, GW170817, and
set bounds on the Jordan-Fierz-Brans-Dicke (JFBD) scalar-tensor theory
of gravity. We note that the \FTI{} approach has already been
  extensively used by LIGO and Virgo data analysts in
  Refs.~\cite{Abbott:2018lct,LIGOScientific:2019fpa,LIGOScientific:2020tif,LIGOScientific:2021sio} and 
also by some of the authors of this manuscript in Ref.~\cite{Sennett:2019bpc}.

This paper is organized as follows. In Sec.~\ref{FTA:sec:Def} we
introduce the \FTI{} method for multipolar waveform models of
compact-object binaries.  After recalling the tenets of Bayesian
  inference in Sec.~\ref{sec:BayesThm}, in Sec.~\ref{sec:results} we
apply the \FTI{} method to synthetic GW signals of BBHs. 
This allows us to discuss the effect of the \FTI{}
  parameterization on the recovery of the BBH properties (excluding
  the GR-deviation parameters) and to study the robustness of the
  \FTI{} method.  In Sec.~\ref{FTA:sec:BransDicke}, we use the \FTI{}
construction on real data, notably the BNS signal GW170817, to set
bounds on the JFBD theory of gravity.  Finally, in
Sec.~\ref{sec:conclusion}, we summarize our main conclusions and also
discuss possible future work. The Appendix \ref{appendix:3.5PNPhasing} 
collects the necessary PN results for the GW phase of BBH with aligned spins.

Henceforth, we use natural units such that the Newton constant $G = 1$ and the speed of light $c = 1$.

\section{The \FTI{} approach}
\label{FTA:sec:Def}

In GR, gravitational signals from quasi-circular BBHs depend on 
the instrinsic parameters $\boldsymbol{\lambda}=\{m_1, m_2,
  \bm{S}_1,\bm{S}_2\}$, where $m_i, \bm{S}_i$ are the masses and spins
  of the compact objects ($i=1,2$), as well as, a set of extrinsic
  parameters $\boldsymbol{\xi} = \{\iota, \varphi_c,\alpha, \delta,
  \psi, d_L, t_c\}$. These are the angular position of the line of
  sight measured in the source frame ($\iota$, $\varphi_c$), the sky
  location of the source in the detector frame $(\alpha,\delta)$, the
  polarization angle $\psi$, the luminosity distance of the source
  $d_L$ and the time of arrival $t_c$. Limiting ourselves to objects
  with non-precessing spins (i.e., spins aligned or antialigned with
  the orbital angular momentum ${\bm{L}}$), the only
    (dimensionless) spin component on which the dynamics, and hence
    the waveform, depends is $\chi_i= \bm{S}_i \cdot
    {\bm{L}}/(|\bm{L}|m_i^2)$. The set of instrinsic parameters
  consequently reduces to four $\boldsymbol{\lambda}=\{m_1, m_2, \chi_{1},
  \chi_{2}\}$. For convenience, we additionally introduce the
    following parameters: the mass ratio $q=m_1/m_2 \geq 1$, the
  symmetric mass ratio $\nu=q/(1+q)^2$, the binary's total mass $M =
  m_1+m_2$, the chirp mass ${\cal M} = \nu^{3/5}\,M$, and the
  effective spin $\chi_{\rm{eff}} = (m_1\chi_1 + m_2\chi_2)/M$. The
  above set of 11 parameters is enough to describe an aligned-spin
    BBH signal. For a binary involving NSs, this set increases by 
    the tidal parameters ($\Lambda_{1,2}$), which encode the NS
    matter equation of state.
  
In GR, the GW signal can be decomposed into a set of modes by projecting the complex linear combination of its plus 
and cross polarizations
\begin{equation}
h(t) \equiv h_{+}(t) -ih_{\times}(t)\,,
\label{eq:GWpol}
\end{equation}
onto spherical harmonics ${}_{-2} Y_{\ell m} $ of spin-weight $-2$~\cite{Ypan2011},
\begin{equation}
h(t;  \bm{\lambda}, \iota, \varphi_c) = \sum^{+\infty}_{\ell=2}\sum_{m=-\ell}^{\ell} {}_{-2}Y_{\ell m} (\iota, \varphi_c) \, h_{\ell m}(t,\bm{\lambda})\,.
\label{eq:hoft_sphericalH}
\end{equation} 
During the inspiral, the GW signals from aligned-spin binaries satisfy a reflection symmetry about its orbital plane, which implies
\begin{equation}
h_{\ell -m}(t) = (-1)^{\ell} h_{\ell m}^{*}(t)\,.
\label{eq:mode_relations}
\end{equation}
where ${}^{*}$ denotes the complex conjugation. As a consequence, we can restrict ourselves to the $m \geq 0$ modes to describe the complete mode-content of aligned-spin inspiral waveforms. Furthermore, for such systems, $\tilde{h}^{R}_{\ell m}(f)$, the Fourier transform of the real part of $h_{\ell m}(t)$ is related to the imaginary part via
\begin{equation}
\tilde{h}^{R}_{\ell m}(f)=-i \tilde{h}^{I}_{\ell m}(f)\,.
\label{eq:mode_relations_freq}
\end{equation}
Using Eq.~(\ref{eq:mode_relations}) and ~(\ref{eq:mode_relations_freq}), the GW polarizations in the frequency domain read (see Appendix C of Ref.~\cite{Mehta:2017jpq} for full derivation)
\begin{subequations}
	\label{eq:hphc_freqdomain_final}
	\begin{align}
	{\tilde h}_+(f) & = \sum^{+\infty}_{\ell=2}\sum_{m=1}^{\ell}\, \Bigr[(-1)^\ell f(\iota)+ 1\Bigr] \, {}_{-2}Y_{\ell m} (\iota, \varphi_c) \, \tilde{h}^{R}_{\ell m}(f)\,, \label{eq:hphc_freqdomain_final_plus} \\ 
	{\tilde h}_\times(f) & = -i \sum^{+\infty}_{\ell=2}\sum_{m=1}^{\ell} \, \Bigr[(-1)^\ell f(\iota)- 1\Bigr] \, {}_{-2}Y_{\ell m} (\iota, \varphi_c) \, \tilde{h}^{R}_{\ell m}(f)\,,
	\end{align}
\end{subequations}
with
\begin{equation}
f(\io)  =  \frac{d_{\,\ell -m}^{\,2}(\io)}{d_{\,\ell m}^{\,2}(\io)}\,, \nonumber
\end{equation}
where $d_{\,\ell m}^{\,2}(\io)$ denote the Wigner functions of weight $-2$~\cite{dfunctions}. Being a complex function, $\tilde{h}^{R}_{\ell m}(f)$ can be written as
\begin{equation}
\tilde{h}^{R}_{\ell m}(f) = A_{\ell m}(f)\, e^{i \, \psi_{\ell m}(f)}. 
\label{eq:GR_modes}
\end{equation}
The construction of a parameterized (or generalized) waveform model
begins with the baseline model in GR (Eq.~(\ref{eq:GR_modes})). During
the quasi-circular, adiabatic inspiral, the frequency-domain phase $\psi_{\ell m}(f)$ can be obtained from PN
theory~\cite{Blanchet:2013haa,Buonanno:2009zt} using the stationary-phase approximation (SPA). In GR it reads
\begin{align}
\psi_{\ell m}^\text{(GR)} (f,\bm{\lambda}) = & \frac{3}{128 \nu v^5} \frac{m}{2}\left[ \sum_{n=0}^{7} \psi_n^{\text{(PN)}}(\bm{\lambda}) v^n   +  \right . \nonumber \\
& \left. \sum_{n=5}^{6} \psi_{n (l)}^{\text{(PN)}}(\bm{\lambda}) v^n \log v \right], \label{FTA:eq:inspPhase}
\end{align}
where 
\begin{equation}
v\equiv \left(  2\pi F M   \right)^{1/3} = \left(  2\pi fM/m   \right)^{1/3} .
\label{eq:orb_freq}
\end{equation}
The quantity $F$ is the orbital frequency which is related to the (Fourier) GW frequency $f$ for a given $(\ell, m)$\nobreakdash-mode through the equation above. The quantities $\psi_n^\text{(PN)}$ and $\psi_{n (l)}^\text{(PN)}$ are the $(n/2)$-PN coefficients~\footnote{The $l$ in $n(l)$ refers to PN coefficients alongside $\log v$ in addition to $v^n$ dependence (see the Appendix \ref{appendix:3.5PNPhasing})}. which depend on the binary parameters. In the 
Appendix~\ref{appendix:3.5PNPhasing}, we provide explicit expressions for the PN coefficients up to 3.5PN order (including 
spin effects), the highest PN order to which the GW phasing is currently known. Note that the logarithmic terms in Eq.~\eqref{FTA:eq:inspPhase} 
arise from tails effects (i.e., terms that depend on the complete past history of the binary).

We generalize the GR waveform model by considering corrections to the phase that take a form similar to the PN expansion
\begin{align}
\delta \psi_{\ell m}(f,\bm{\lambda}; \delta \hat{\varphi}_n, \delta \hat{\varphi}_{n(l)}) & = \frac{3}{128 \nu v^5} \frac{m}{2} \left[ \sum_{n=-2}^{7} \delta \psi_n(\bm{\lambda}, \delta \hat{\varphi}_n) v^n \right. \nonumber \\
& \left. +\sum_{n=5}^{6} \delta \psi_{n (l)}(\bm{\lambda}, \delta \hat{\varphi}_{n(l)}) v^n \log v \right],
\label{FTA:eq:deltaPsi}
\end{align}
where $\delta \psi_n$ and $\delta \psi_{n (l)}$ are deviations to the $(n/2)$-PN phase coefficients $\psi_n^\text{(PN)}$ and $\psi_{n (l)}^\text{(PN)}$ defined above. These correction terms depend, in addition to $\bm{\lambda}$, also 
on the corresponding deviation parameter $\delta \hat{\varphi}_n$ or $\delta \hat{\varphi}_{n (l)}$, respectively.
We include possible deviations at ``pre-Newtonian'' orders ($n<0$) as these are predicted in some alternative theories of gravity.
{In particular the emission of dipole radiation (discussed in Sec.~\ref{sec:Signature}) leads to a nonvanishing deviation at $n=-2$.
A solitary deviation at $n=-1$ is less well motivated and not discussed in this work.}
Parameterized deviations of this form can be mapped onto the predictions of any hypothetical theory of gravity provided that (i) the theory admits a weak-field, slow-velocity PN expansion as in GR, and (ii) the deviations from GR are parametrically smaller than the PN-expansion parameter $v^2$. Note that this excludes theories that admit non-perturbative phenomena like dynamical scalarization, for which the naive PN expansion in Eq.~\eqref{FTA:eq:deltaPsi} breaks down~\cite{Palenzuela:2013hsa,Sennett:2016rwa}, {and other methods are needed to observe such effects (see, e.g., Ref. \cite{Edelman:2020aqj})}. Because GW detectors are more sensitive to the evolution of the signal's phase than its amplitude, considering also the computational cost of introducing 
  several free parameters, we neglect deviations in the mode amplitudes $A_{\ell m}$.
  
  Despite the generality of Eq.~\eqref{FTA:eq:deltaPsi}, the moderate
  signal-to-noise ratios (SNRs) of most LIGO-Virgo observations, so
  far, do not allow us to place meaningful bounds on multiple
  deviation parameters concurrently.  Hence, we vary one parameter at
  a time, keeping the rest fixed at their nominal GR prediction,
  which is zero. This assumption is validated by
  investigations~\cite{Sampson:2013lpa,Meidam:2017dgf,Perkins:2021mhb, Perkins:2022fhr} that conclude that a signal
  containing deviations at several PN orders is likely to lead to a
  non-zero deviation measurement using a model with only a single
  deviation parameter. On the other hand, a recent
  work~\cite{Saleem:2021nsb} following a principle component
    analysis identified certain combinations of these deviation
    parameters with the tightest constraints; in fact, the two
    dominant principle components rather capture the essence of the
    full multi-parameter test. 

In this paper, we assume that each deviation parameter represents a fractional deviation to
the corresponding PN coefficient in GR and for this reason, we
  also refer  to the deviation parameters as non-GR parameters, 
\begin{subequations}
	\begin{align}
	\delta \psi_n(\bm{\lambda}, \delta \hat{\varphi}_n) & \equiv \delta \hat{\varphi}_n \psi_n^{(\text{PN})}(\bm{\lambda}), \\
	\delta \psi_{n (l)}(\bm{\lambda}, \delta \hat{\varphi}_{n (l)}) & \equiv \delta \hat{\varphi}_{n (l)} \psi_{n(l)}^{(\text{PN})}(\bm{\lambda}).
	\end{align}
  \label{eq:deltapsi}%
\end{subequations}
We handle PN orders for which GR coefficients vanish
slightly differently (i.e., for $n=-2$, $-1$, $1$). For such cases, we let
$\varphi_n$ represent an \emph{absolute} deviation at
that order instead.

While Eq.~\eqref{FTA:eq:deltaPsi} unambiguously details how to
generalize GR waveforms containing only the inspiral, additional care
must be taken for waveforms that contain later portions of the GW
signal, such as the merger-ringdown.  For the \FTI{} approach, we 
require that the parameterized deviations satisfy the following properties
\begin{enumerate}
	\item {The early-inspiral (low-frequency) waveform has a phase $\psi_{\ell m}(f;\bm{\lambda}; \delta \hat{\varphi}_n,\delta \hat{\varphi}_{n (l)})=\psi_{\ell m}^\text{(GR)}(f;\bm{\lambda}) + \delta \psi_{\ell m}(f;\bm{\lambda};\delta \hat{\varphi}_n,\delta \hat{\varphi}_{n (l)})$, where $\delta \psi_{\ell m}$ takes the form of Eq.~\eqref{FTA:eq:deltaPsi}.}
	\item The post-inspiral (high-frequency) waveform has a phase $\psi_{\ell m}(f;\bm{\lambda}; \delta \hat{\varphi}_n,\delta \hat{\varphi}_{n (l)}) = \psi_{\ell m}^{(\text{GR})}(f;\bm{\lambda})$ that exactly reproduces the underlying GR polarizations (Eq.~\eqref{eq:hphc_freqdomain_final}) up to some constant shift which represents the total dephasing from the GR polarization accumulated over the inspiral.
	\item The waveform polarizations are $C^2$ smooth over all frequencies.
\end{enumerate}
Using Eq.~\eqref{eq:deltapsi}, the $\delta \psi_{\ell m}(f;\bm{\lambda};\delta \hat{\varphi}_n,\delta \hat{\varphi}_{n (l)})$ in Eq.~\eqref{FTA:eq:deltaPsi} read
\begin{align}
\delta \psi_{\ell m}(f,\bm{\lambda};\delta \hat{\varphi}_n,\delta \hat{\varphi}_{n (l)}) \equiv &\frac{3}{128 \nu v^5} \frac{m}{2}\left[ \sum_{n=-2}^{7} \psi_n^{\text{(PN)}}(\bm{\lambda}) \delta \hat{\varphi}_n v^n \right. 
\nonumber \\ 
& \left.  +\sum_{n=5}^{6} \psi_{n (l)}^{\text{(PN)}}(\bm{\lambda}) \delta \hat{\varphi}_{n (l)} v^n \log v \right]. \label{eq:deltaPN}
\end{align}
To smoothly apply these corrections over only the inspiral, we use a tapering function $W(f;v^\text{tape}, \Delta v^\text{tape})$ given by
\begin{align}
W(f;v^\text{tape}, \Delta{v}^\text{tape}) \equiv \left[1+\exp \left(\frac{v-v^\text{tape}}{\Delta v^\text{tape}}\right)\right]^{-1}, \label{eq:wind_func}
\end{align}
which smoothly transitions between one and zero around $v^\text{tape}$ over
the range of $\sim \Delta v^\text{tape}$, where $v$ is defined in
Eq.~\eqref{eq:orb_freq}. We construct the total phase correction for a given
$(\ell, m)$\nobreakdash-mode by combining this windowing function with the
second derivative with respect to frequency of $\delta \psi_{\ell m}$, which we denote as $\psi_{\ell m}^{\prime\prime}(f'')$,
and re-integrating with appropriate integration constants to ensure
$C^2$ smoothness. In summary, we use
\begin{align}
& \delta \psi_{\ell m}(f;\bm{\lambda};\delta \hat{\varphi}_n,\delta \hat{\varphi}_{n (l)};v^\text{tape}, \Delta{v}^\text{tape})  = \nonumber \\
& \;\; \int_{f_{\ell m}^\text{ref}}^f df^\prime \int_{f_{22}^\text{peak}}^{f^\prime} d f^{\prime \prime} \delta \psi_{\ell m}^{\prime\prime}(f^{\prime \prime},\bm{\lambda};\delta \hat{\varphi}_n,\delta \hat{\varphi}_{n (l)}) \nonumber \\
& \;\; \times W(f^{\prime \prime};v^\text{tape}, \Delta{v}^\text{tape}), \label{eq:phase_corr}
\end{align}
where $f_{\ell m}^\text{ref}$ $ (\equiv m f_{22}^\text{ref} /2)$ is the reference frequency at
which the phase of the ($\ell m$)-mode  vanishes. This choice ensures that the definition of the reference frequency does not change when we add these corrections to the GR waveform. The second integration boundary is fixed by requiring that the first derivative of
$\delta \psi_{\ell m}(f)$ goes to zero at the frequency of the $(2,2)$-mode's peak
$f_{22}^{\text{peak}}$ (see Eq.~(A8) of Ref.~\cite{Bohe:2016gbl}). This requirement ensures that the alignment between the GR waveform and the modified GR waveform in the time-domain remains the same. 

\begin{figure*}[htb!]
	\begin{center}
		\includegraphics[width=0.49\textwidth]{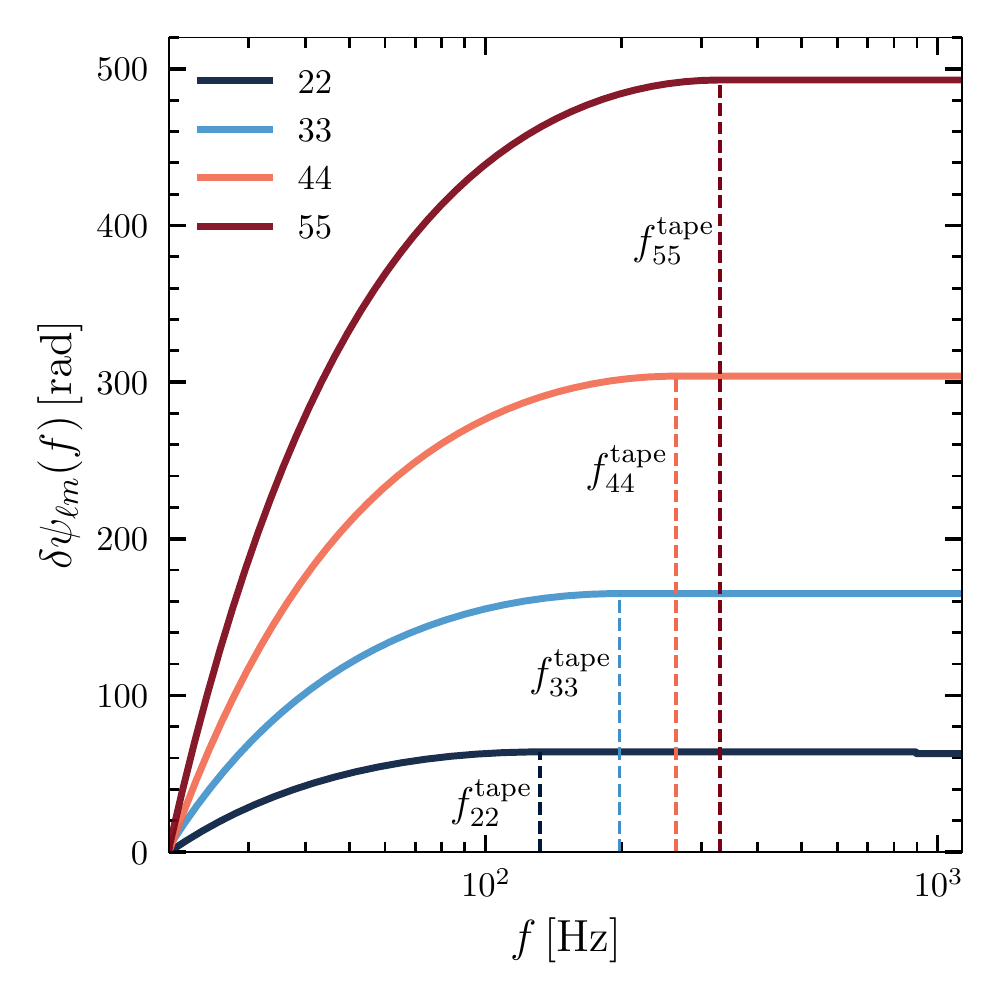}
		\includegraphics[width=0.49\textwidth]{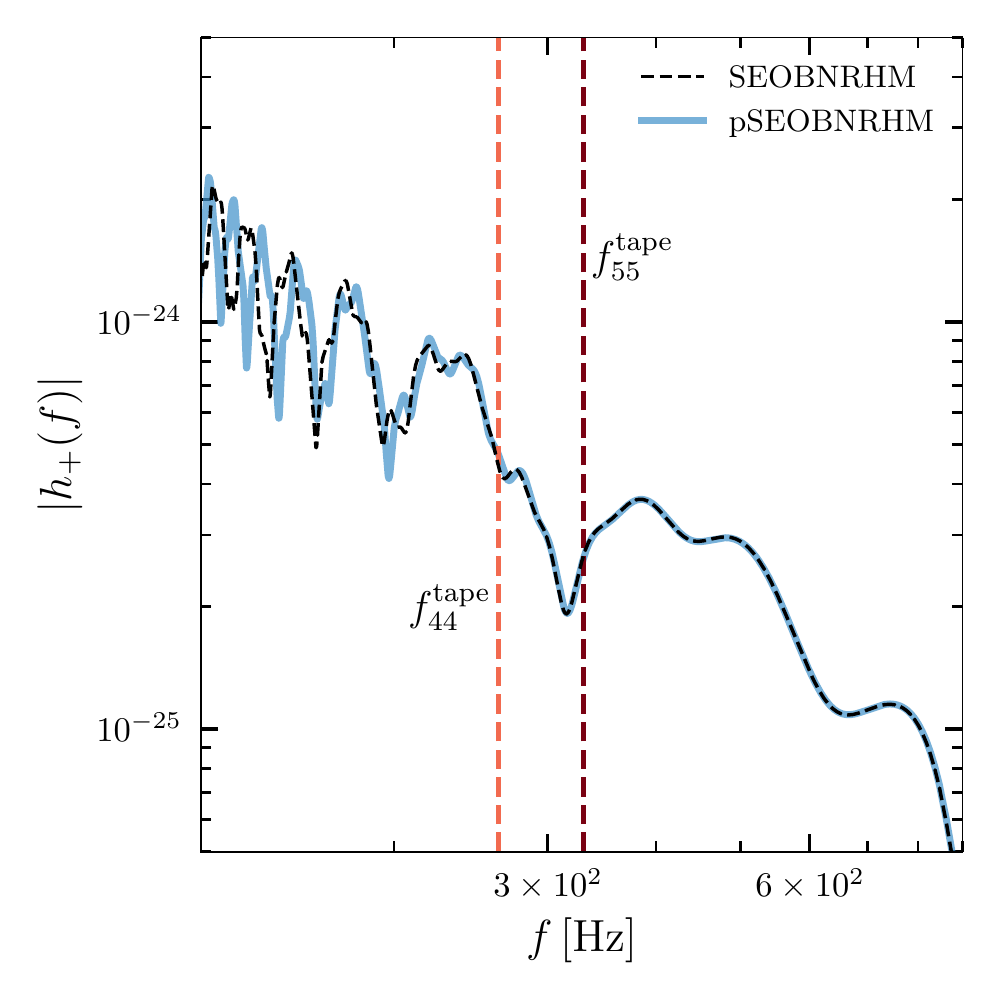}
	\end{center}
	\caption{\textit{Left panel}: The phase corrections for the different frequency-domain modes for the parameters of a GW190814-like event~\cite{GW190814} (see Table~\ref{table:nedian_params}). The numerical value of the $(2,1)$-mode is relatively small and hence it is not shown in the plot for the sake of clarity. We use as 
tapering frequency $f^{\text{tape}}_{22} = 0.35 f_{22}^{\mathrm{peak}}$. The vertical dashed lines denote the tapering frequencies $f^{\text{tape}}_{\ell m}$ for each mode. The phase correction for each mode becomes constant for $f > f^{\text{tape}}_{\ell m}$ where they are tapered off. \textit{Right panel}: The amplitude of the plus polarization of the \SEOBNRHM{} and \pSEOBNRHM{} waveforms. The amplitude of the \pSEOBNRHM{} waveform returns to the GR (\SEOBNRHM{}) one in the post-inspiral regime  ($f > f^{\rm{tape}}_{44}$).} 
	\label{fig:fta_example_fd}
\end{figure*}

Let us elaborate more on the tapering parameters $v^{\rm{tape}} $ and
$\Delta{v}^\text{tape}$ that enter the final expression for the phase
corrections~\eqref{eq:phase_corr}. Using Eq.~\eqref{eq:orb_freq}, the
parameter $v^{\rm{tape}} $ can be equivalently specified by the
orbital frequency $F^{\mathrm{tape}}$ at which the corrections are
tapered off. Note that while the orbital tapering frequency
$F^\text{tape}$ is the same for all modes, the tapering frequency in
Fourier domain $f^{\text{tape}}_{\ell m} = m F^{\text{tape}}$ depends
on the mode. In the following, we fix the tapering frequency by
specifying $f^{\text{tape}}_{22}$ as a fraction of
$f_{22}^{\rm{peak}}$, say $f^{\text{tape}}_{22} = \alpha
f_{22}^{\rm{peak}}$ with a constant $\alpha$ of order unity, and
$v^\text{tape} = (\pi f_{22}^{\rm{tape}} M)^{1/3}$. Furthermore,
instead of specifying the parameter $\Delta{v}^\text{tape}$ directly,
we find it more useful to fix the number of GW cycles over which the
window function defined in Eq.~\eqref{eq:wind_func} switches its value
from 0 to 1. The number of GW cycles $\mathcal{N}_{\rm GW}$ between
the GW frequencies $f_1$ and $f_2$, or respectively $v_1$ and $v_2$
defined by Eq.~\eqref{eq:orb_freq}, can be estimated as
\begin{equation}
\mathcal{N}_{\rm GW} = \int_{f_1}^{f_2} f (t) dt = \int_{f_1}^{f_2} df \dfrac{f}{\dot{f}} = \dfrac{1}{32\pi \nu} (v_1^{-5}-v_{2}^{-5}), 
\end{equation}
where we make use of the leading-order PN expressions for $f(t)$ and $\dot{f}(t)$ in the last equality.
Now, choosing $v_{1,2}=v^{\mathrm{tape}}\mp\Delta v^{\mathrm{tape}}/2$, we solve for the small $\Delta{v}^\text{tape} \ll v^\text{tape} $,
\begin{equation}
  \Delta{v}^\text{tape} = \frac{128 \nu}{3} \pi (v^{\rm{tape}})^6 \,\Gamma \, \mathcal{N}_{\rm GW} .
\end{equation}
with the estimate $\Gamma \equiv \Gamma_\text{PN} = 3/20$ from PN theory. However, since we are going to apply the tapering close to the merger of the binary, where the PN approximation is not applicable, we treat $\Gamma$ as a phenomenological fudge factor and choose $\Gamma = 1/50$. The crucial input from PN theory is the functional dependence of $\Delta{v}^\text{tape}$ on $\nu$, $v^{\rm{tape}}$ and $\mathcal{N}_{\rm GW}$. The choice of the parameters $f^{\text{tape}}_{22}$ and $\mathcal{N}_{\rm GW}$ is completely phenomenological, and should be made to optimize the null test. Previous analyses on the LIGO-Virgo events reported in Refs.~\cite{LIGOScientific:2020tif, GW190412, GW190814,  LIGOScientific:2021sio} used $f^{\text{tape}}_{22} =  0.35 f_{22}^{\rm{peak}}$ and $\mathcal{N}_{\rm GW}=1$. While we also employ these choices for the studies here, we devote Sec.~\ref{robustness} to investigate how results are effected when varying, in particular, the tapering frequency $f^{\text{tape}}_{22}$. Changing $\mathcal{N}_{\rm GW}$ in the range $0.8\mbox{--}3$, instead, does not affect the results significantly.

At low frequencies, any GR description of the modes of the GW-phase reduces to the one of Eq.~(\ref{FTA:eq:inspPhase}). Thus, we can apply the 
method outlined above to inspiraling PN models and also to inspiral-merger-ringdown GR models  (if they are available in time domain we first perform a Fourier transform). Our method can treat the deviation coefficients in Eqs.~(\ref{FTA:eq:deltaPsi}) and (\ref{eq:deltapsi}) either 
as free parameters (see Sec.~\ref{sec:results}) or identify them to the ones predicted in specific alternative theories of gravity 
(see Sec.~\ref{FTA:sec:BransDicke}), provided the latter have perturbative deviations from GR 
and are represented by PN-like coefficients. This is the eponymous flexibility of the \FTI{} approach.

In this work, we apply the \FTI{} approach to the multipolar, aligned-spin effective-one-body (EOB) waveform model 
\SEOBNRHM{}~\cite{Bohe:2016gbl,Cotesta:2020qhw} for BBHs, the aligned-spin EOB model 
\SEOBNRT{}~\cite{Bohe:2016gbl,Hinderer:2016eia,Lackey:2018zvw} for BNSs, and for some studies, the aligned-spin 
inspiral-merger-ringdown phenomenological model \PhenomX{} for BBHs~\cite{Pratten:2020fqn}~\footnote{In the LIGO Algorithm
  Library (LAL) the technical names of these waveform models are 
{\tt SEOBNRv4HM}$\_{\rm ROM}$, {\tt SEOBNRv4T}$\_{\rm surrogate}$, and {\tt IMRPhenomXAS}, respectively.}. The \SEOBNR{}, \SEOBNRT{} and 
\PhenomX{} waveforms contain the $(\ell, m) =(2,2)$ (dominant) mode, while the \SEOBNRHM{} waveforms include four additional 
sub-dominant modes, $(\ell, m) = (2,1)$, $(3,3)$, $(4,4)$, and $(5,5)$. We denote the waveforms to which we apply the non-GR phase corrections (\ref{eq:deltapsi}) 
\textit{parameterized} waveforms and refer to them as \pSEOBNRHM{}, \pSEOBNRT{}\ and \pPhenomX{}.

\begin{figure*}[htb!]
	\begin{center}
		\includegraphics[width=1\textwidth]{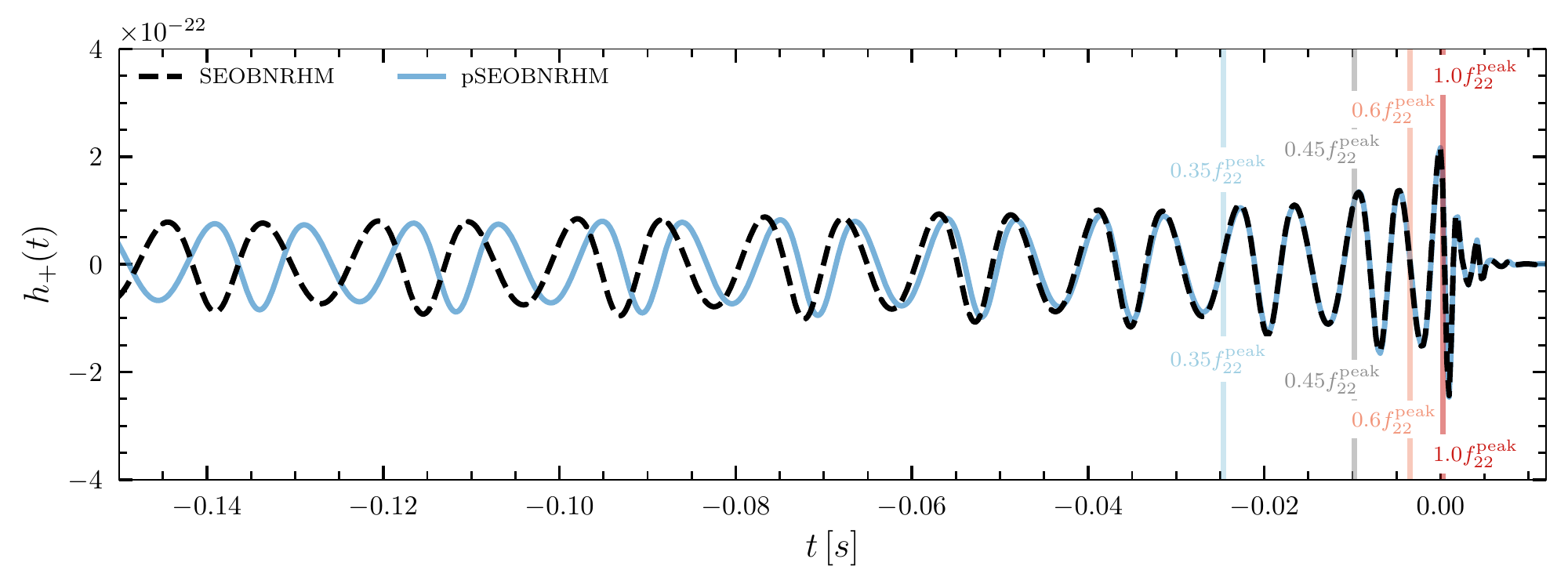}
	\end{center}
	\caption{Plus polarization of the GR (\SEOBNRHM{}) and \pSEOBNRHM{} waveforms shown in Fig.~\ref{fig:fta_example_fd}. The waveforms are aligned at their peaks, $t=0$. The solid vertical lines denote the times corresponding to the different tapering frequencies that we employ in Sec.~\ref{robustness}. Here, we use as tapering frequency in the \pSEOBNRHM{} waveform $f^{\text{tape}}_{22} = 0.35 f_{22}^{\mathrm{peak}}$. The \SEOBNRHM{} and the \pSEOBNRHM{} waveforms match very well in the post-tapering region (i.e., for $t\geq t^{\rm{tape}}$ where $2F(t^{\rm{tape}})=0.35 f_{22}^{\mathrm{peak}}$).}
	\label{fig:fta_example_td}
\end{figure*}

We end this section with an illustration of the tapering using the \pSEOBNRHM{} model. 
We choose a binary with parameters that correspond to the median of the GW190814~\cite{GW190814} event, as listed in Table~\ref{table:nedian_params}, and
generate two \pSEOBNRHM{} waveforms corresponding to a
  deviation parameter of $\delta \hat{\varphi}_{2}=0.5$ and its GR
  limit $\delta \hat{\varphi}_{2}=0$ (i.e., identical to \SEOBNRHM{}).
Figure~\ref{fig:fta_example_fd} shows contributions of the phase
corrections $\delta \psi_{\ell m}^\text{(PN)}(f)$ to the different
($\ell, m$)-modes (left panel), as well as the amplitude of the plus
polarizations after combining all the modes (via
Eq.~\eqref{eq:hphc_freqdomain_final_plus}) for the \SEOBNRHM{} and
\pSEOBNRHM{} waveforms (right panel). The vertical dashed lines
mark the tapering frequency $f^{\text{tape}}_{\ell m}$ associated with
the different modes. The left panel highlights how the correction to
the phase for each mode, $\delta \psi_{\ell m}(f)$, approaches a
constant value for $f > f^{\text{tape}}_{\ell m}$ where the
correction is tapered off. Consistent with this, the right panel shows
how the amplitude of the parameterized waveform (\pSEOBNRHM{})
returns to the GR limit (\SEOBNRHM{}) for $f >
  f_{44}^{\rm{tape}}$~\footnote{Note that, even though the
  $(5,5)$-mode is included in the \SEOBNRHM{} waveform, it does
  not contribute much in this case.} We note that the parameterized
waveform returns to the GR one when the overall phase difference, as
  illustrated in the left panel, becomes constant and is hence
  absorbed into the coalescence phase. In other words, the
coalescence phase recovered using the parameterized waveforms and GR
waveforms from the parameter-estimation  studies is, in general,
not expected to agree. In Fig.~\ref{fig:fta_example_td}, we show the plus 
polarization of the parameterized and GR waveforms in time domain. The time-domain
waveforms are generated from their frequency-domain counterparts via
inverse Fourier transformation. The waveforms are aligned at the peak
of their amplitudes. The vertical lines indicate the times
  corresponding to the tapering frequencies (i.e., $F (t^{\rm{tape}})
  \equiv F^{\rm{tape}}$), where the quantity $2F(t)$ is computed from
  the first derivative of the (2,2)-mode's phase. For this figure, the tapering frequency used is
  $f^{\text{tape}}_{22} = 0.35 f_{22}^{\rm{tape}}=2F^{\rm{tape}} $, as
  stated before. For later analyses, we also indicate in the figure 
other choices of the tapering frequency. In the time-domain plot, it
is visually clearer to see that the \pSEOBNRHM{} waveform
reduces to \SEOBNRHM{} post-tapering (i.e., for $t \geq
t^{\rm{tape}}$), while displaying a significant mismatch before
the tapering (in the inspiral regime).

\section{Bayesian Inference}
\label{sec:BayesThm}

We can now use the parameterized waveforms introduced under the \FTI{}
approach in the previous section to perform tests of GR with
LIGO-Virgo observations. The first step involves the measurement or
\emph{inference} of the waveform parameters, given the observed
GW data. For this, we use a Bayesian approach in this paper. Given the
hypothesis $\mathcal{H}$ that our data $d = h(\boldsymbol{\theta})+n$
consists of detector noise $n$ and a single GW signal, which can
be accurately described by our waveforms $h(\boldsymbol{\theta})$
depending on parameters $\boldsymbol{\theta}=\{ \boldsymbol{\lambda},
\boldsymbol{\xi}, \delta \hat{\varphi}_{n}, \delta
\hat{\varphi}_{n(l)} \}$, the Bayes' theorem states that
\begin{equation}
P(\boldsymbol{\theta}|d, \mathcal{H}) = \dfrac{\mathcal{L}(d|\boldsymbol{\theta}, \mathcal{H}) p(\boldsymbol{\theta}|\mathcal{H})}{E(d|\mathcal{H}) } .
\label{eq:Bayes_thm}
\end{equation}
Here, $P(\boldsymbol{\theta}|d, \mathcal{H})$ is the posterior probability distribution of the parameters $\boldsymbol{\theta}$, and $p(\boldsymbol{\theta}|\mathcal{H})$ is the prior probability distribution. The quantity $E(d|\mathcal{H})$ is called the \textit{evidence} of the hypothesis $\mathcal{H}$, which is just the normalization constant of the  posterior probability distribution $P(\boldsymbol{\theta}|d, \mathcal{H})$. 
The quantity $\mathcal{L}(d|\boldsymbol{\theta}, \mathcal{H})$ denotes the likelihood of obtaining the data $d = h(\boldsymbol{\theta}) + n$ given that the parameters of the waveform are $\boldsymbol{\theta}$ under the hypothesis $\mathcal{H}$, the latter entailing the waveform and noise models. It is computed using
\begin{equation}\label{eq:likelihood}
\mathcal{L}(d|\boldsymbol{\theta}, \mathcal{H}) \propto \exp\left[-\dfrac{1}{2} \langle d-h(\boldsymbol{\theta}) | d-h(\boldsymbol{\theta}) \rangle \right],
\end{equation} 
where $\langle .|. \rangle$ is the noise-weighted inner product,
\begin{equation}
\langle a| b \rangle = 2\int_{f_{\rm{low}}}^{f_{\rm{high}}} \dfrac{\tilde{a}^{*}(f) \tilde{b}(f) + \tilde{a}(f) \tilde{b}^{*}(f)}{S_{n}(f)} df\,.
\end{equation}
Equation~\eqref{eq:likelihood} simply encodes our hypothesis of a
stationary Gaussian noise model, with the power spectral-density
(PSD) of the GW detector noise denoted by $S_n(f)$. We work with the
three-detector network consisting of the two LIGO and the Virgo
detectors. The quantities $f_{\rm{high}}$ and $f_{\rm{low}}$ denote
the maximum and minimum frequencies, respectively, that enter in the
likelihood estimation. The precise values of these quantities are
dictated by the sensitivity bandwidth of the GW detectors. We will
state them explicitly wherever required.

The prior probability distributions chosen for our analyses in this
paper are identical to Ref.~\cite{Veitch:2014wba}. More specifically, 
they are uniform in the component masses, isotropic in spin orientations and uniform in their magnitudes between $[0,0.99]$, uniform in the
Euclidean volume for the luminosity distance, and isotropic in the sky
location and binary orientation. We also choose a flat prior on our
non-GR parmeters, $\delta \hat{\varphi}_{n}$, $\delta
\hat{\varphi}_{n(l)}$. For the GW170817 analysis in
Sec.~\ref{FTA:sec:BransDicke}, we will introduce
  additional parameters and their priors below. With these prior
distributions, we proceed to stochastically sample the parameter space
using a Markov-Chain Monte Carlo (MCMC) algorithm provided in the
$\tt{LALINFERENCE}$ code~\cite{Veitch:2014wba}. The one-dimensional
posterior distribution of a specific parameter (demonstrated,
  e.g., in Fig.~\ref{fig:violin_GW190814}) is then computed by simply
marginalizing $P(\boldsymbol{\theta}|d, \mathcal{H})$ over the
\emph{nuisance} parameters.

We now utilize the Bayesian parameter inference outlined above to demonstrate 
the \FTI{} approach in two specific cases: 
in the context of a theory-agnostic test with the BBH GW signals GW190412 
and GW190814 (Sec.~\ref{sec:results}), and a theory-specific test with the 
BNS signal GW170817 (Sec.~\ref{FTA:sec:BransDicke}).

\section{Application of the \FTI{} approach to binary black holes}
\label{sec:results}

The first of our two goals in this paper is to show the pliability of
the \FTI{} approach in performing theory-agnositc tests of GR, where
one checks for the (dis)agreement between estimates of non-GR
parameters with their GR predictions. In this section, we stress-test
the \FTI{} approach on GW signals, and explore its robustness against
different assumptions made by different families of waveforms. We also
investigate possible systematic biases when the underlying waveform
model contains missing physics, for example, an absence of information
due to higher modes (HMs). Finally, we scrutinize how ``flexible" the \FTI{}
approach really is in its choice of internal settings like the
tapering frequency. An incomplete model or an incorrect internal
setting can potentially flag a violation of GR, and needs to be
accounted for while performing theory-agnositc tests of GR.

These investigations also illuminate the robustness of parameterized
theory-agnostic inspiral tests in general, which is more
straightforward in an approach like \FTI{} that strives to expose all
flexibility (or arbitrariness) of the test.  Comparing this to the
\texttt{TIGER} infrastructure~\cite{Li:2011cg,Agathos:2013upa}, we
note that \texttt{TIGER} is designed and implemented within a
particular family of waveform models, with a specific prescription for
the transition to the merger-ringdown phase being essentially
inherited from the underlying GR waveform.

\subsection{\FTI{} results with synthetic signals: GW190412-like and GW190814-like}

\begin{table}
	\centering
	\caption{Median values and symmetric 90\% credible intervals for the source-frame parameters of the  GW190814 \cite{GW190814} and GW190412 \cite{GW190412} signals observed by LIGO and Virgo. We indicate the source-frame masses with the superscript $s$, to distinguish them from 
the detector-frame masses, $m_i = (1+z)\,m_i^s$, used in the rest of the paper. We use the median values for our synthetic signals: GW190814\textendash like and GW190412\textendash like signals.}
	\begin{tabular}{ccc} 
		~Parameter~ & ~GW190814~ & ~GW190412~ \\
		\hline \rule{0pt}{1.1\normalbaselineskip}
		$m^s_{1}/M_\odot$ & $23.2_{-1.0}^{+1.1}$ & $30.1_{-5.3}^{+4.6}$ \\[0.2cm]
		$m^s_{2}/M_\odot$  & $2.59^{+0.08}_{-0.09}$ & $8.3_{-0.9}^{+1.6}$ \\[0.2cm]
		$1/q$ & $0.112_{-0.009}^{+0.008}$ & $0.28_{-0.07}^{+0.12}$ \\[0.2cm]
		$\iota$ (rad) & $0.8_{-0.2}^{+0.3}$ & $0.71_{-0.24}^{+0.31}$ \\[0.2cm]
		$z$ & $0.053_{-0.010}^{+0.009}$ & $0.15_{-0.03}^{+0.03}$ \\[0.2cm]
		$\rm{SNR}$ & $25.0_{-0.2}^{+0.1}$ &  $19.1_{-0.3}^{+0.1}$ \\[0.1cm]
	\end{tabular}
	\label{table:nedian_params}
\end{table}

\begin{figure*}[!htbp]
	\begin{center}
		\includegraphics[width=1\textwidth]{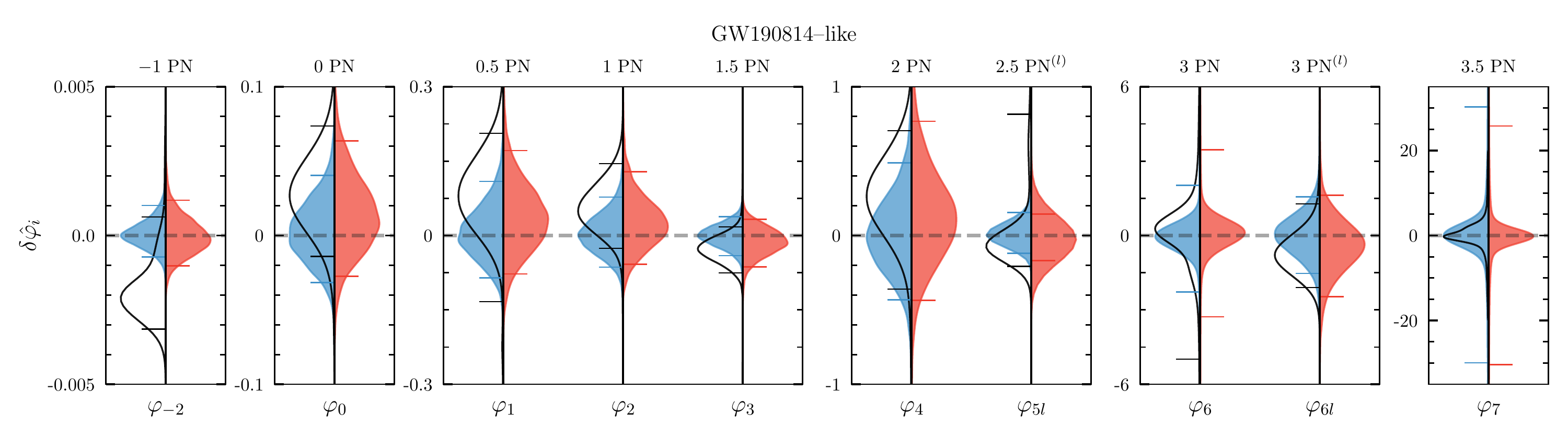}
		\includegraphics[width=1\textwidth]{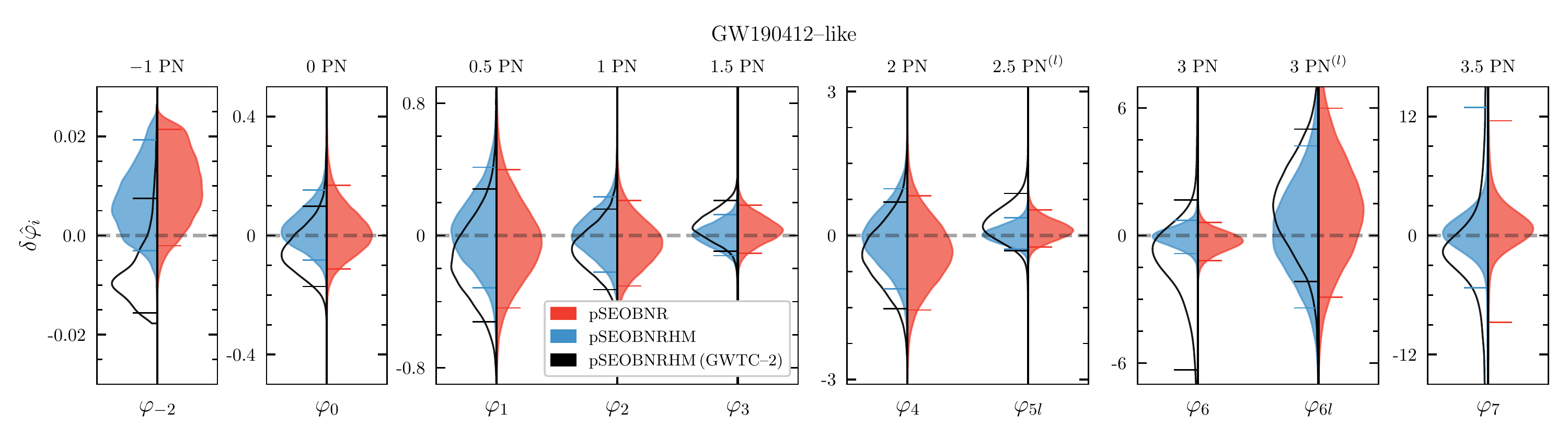}
	\end{center}
  \caption{Filled histograms for the posterior distributions of the deviation parameters $\delta \hat \varphi_{i}$, $\delta \hat \varphi_{i(l)}$ appearing in Eq.~\eqref{eq:deltaPN} when a GR simulated signal using the \pSEOBNRHM\, waveform is analyzed with the parameterized waveforms \pSEOBNRHM\, (blue) and 
\pSEOBNR\, (orange) (top panel: GW190814\textendash like parameters; bottom panel: GW190412-like parameters). The inclusion of HMs improves both the precision and the accuracy of the measurement of the deviation parameters. Also shown by the black unfilled histograms are the results obtained from the actual GW190814 and GW190412 events when analyzed with the \pSEOBNRHM\,, as first presented in Ref.~\cite{LIGOScientific:2020tif} (but here without reweighing against the priors). The small symmetrical horizontal lines in each panel represent the 90\% credible intervals.}
	\label{fig:violin_GW190814}
\end{figure*}

In this section, we explore the effect on theory-agnostic tests of 
varying the physical content in our waveform model, for instance, the
impact of HMs. We restrict ourselves to GW signals
from BBH mergers, specifically, the high-mass ratio GW190412 and
GW190814 BBH signals, which show non-negligible evidence for the
presence of HMs~\cite{GW190412, GW190814}. We note that the \FTI{}
analysis with HMs has been already performed by the LVC on these two events in
Ref.~\cite{LIGOScientific:2020tif} (see Appendix C therein). Let us
  recapitulate those results here, summarized by the unfilled black
  histograms in Fig.~\ref{fig:violin_GW190814}.  However, note that in
  LVC publications, the \FTI{} results are typically transformed and
  reweighed to match the normalization of the \TIGER{} inspiral
  parameters.  This permits comparisons between the two approaches,
  generally finding good agreement for the GR constraints (see also
  Sec.~V of Ref.~\cite{LIGOScientific:2020tif} for details).
  Throughout this work, we present the \FTI{} results without such
  reweighing.

 
\begin{figure*}[htb!]
	\begin{center}
		\includegraphics[width=0.49\textwidth]{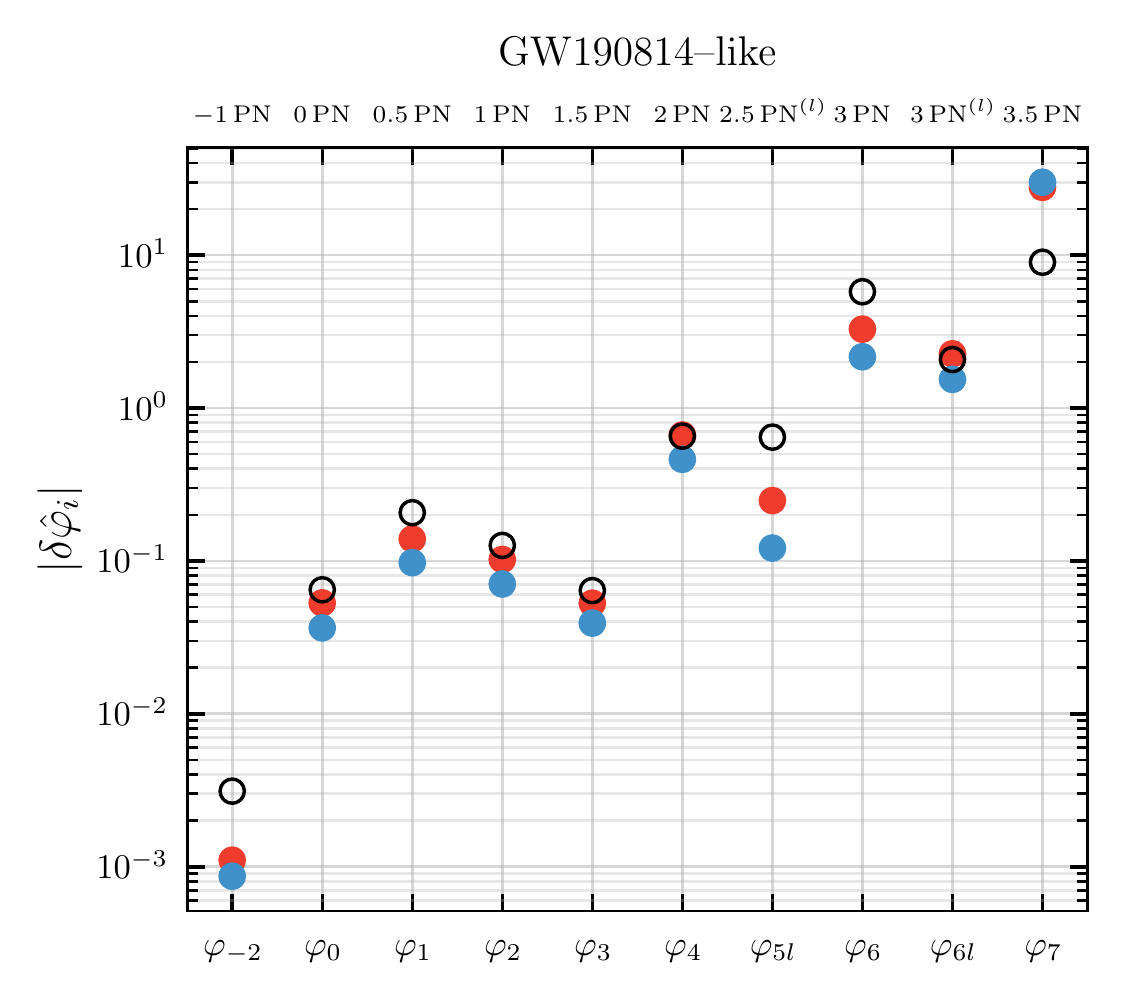}
		\includegraphics[width=0.49\textwidth]{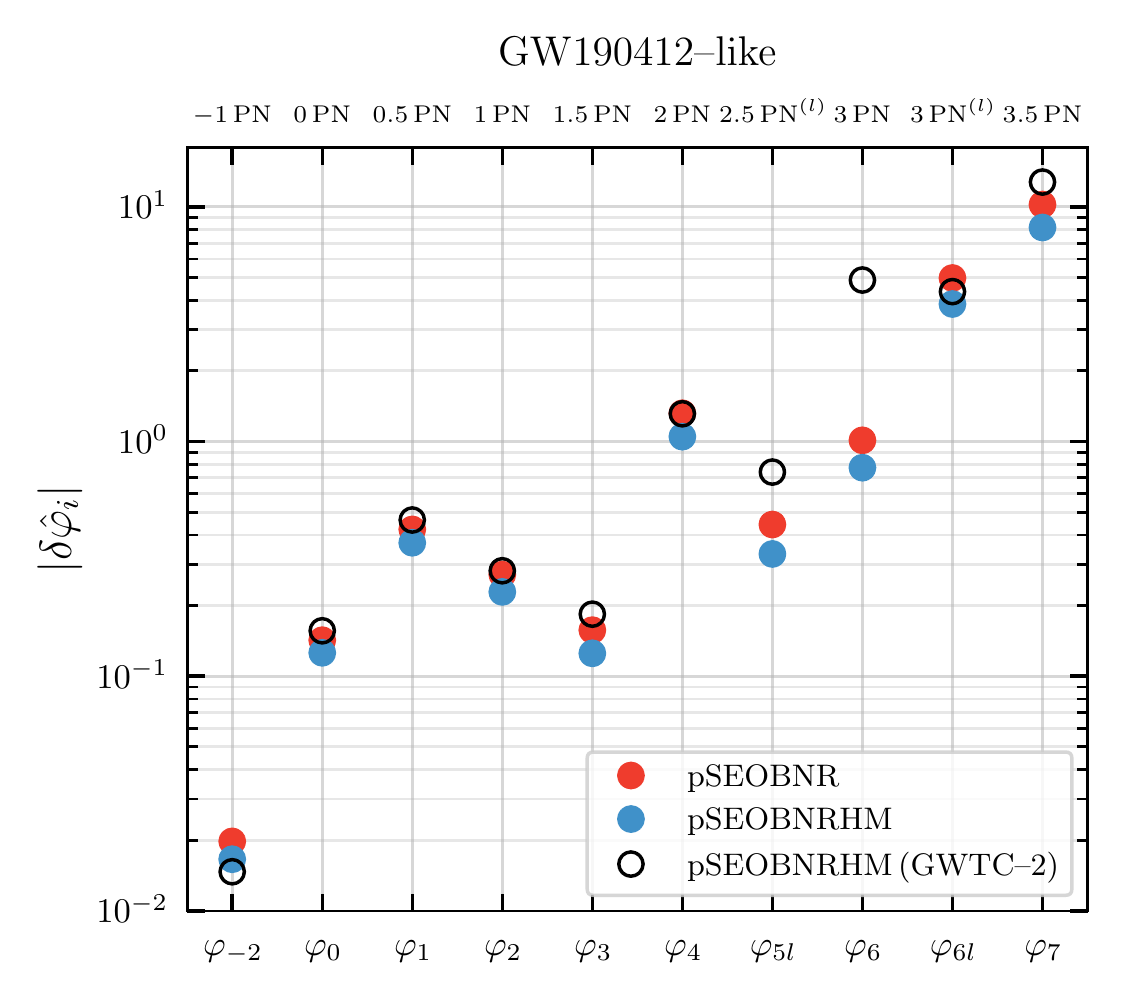}
	\end{center}
  \caption{The 90\% credible intervals of the posterior distributions of the deviation parameters in Fig.~\ref{fig:violin_GW190814} (i.e., for the simulated GW190814\textendash like (left panel) and GW190412\textendash like (right panel) signals.) The HM waveforms, \pSEOBNRHM\,, provide slightly better bounds. The open circles are the results from the actual events GW190814 and GW190412~\cite{LIGOScientific:2020tif} (but here without reweighing against the priors).}
	\label{fig:pn_bounds}
\end{figure*}

\begin{figure}
	\begin{center}
		\includegraphics[width=0.49\textwidth]{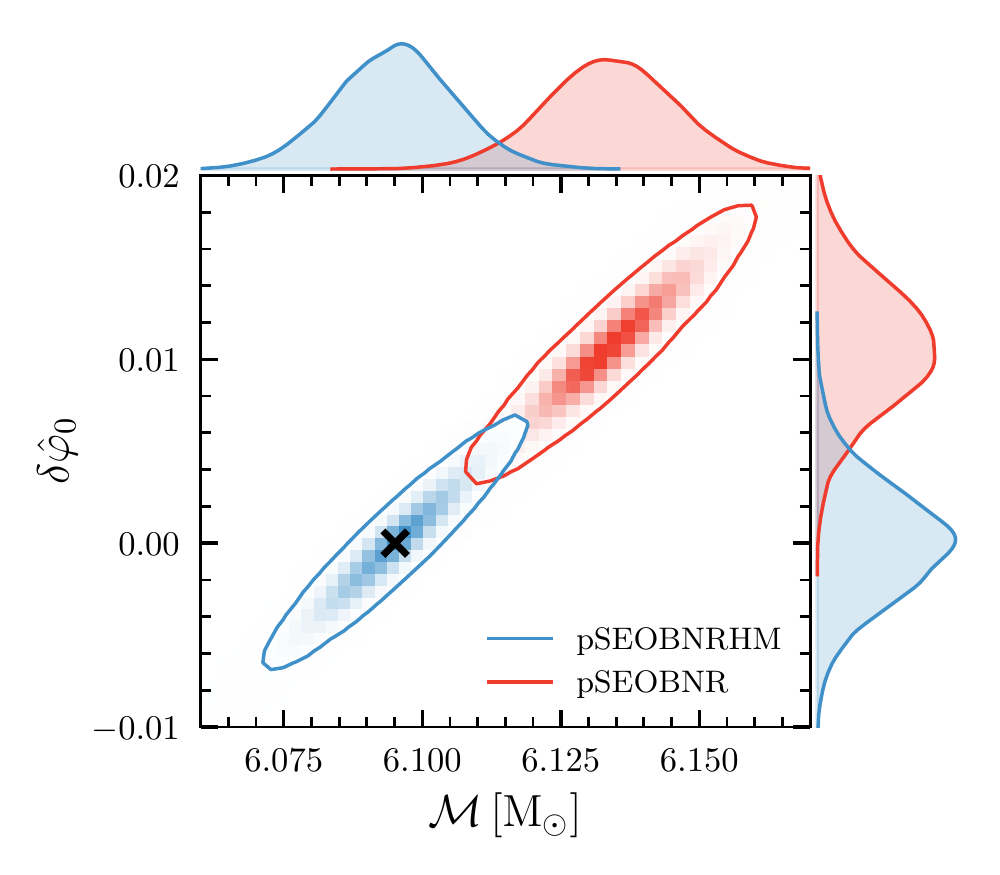}
	\end{center}
	\caption{The same analysis as in Fig.~\ref{fig:violin_GW190814} ($\mathrm{SNR}=25$) for the deviation parameter $\delta \hat{\varphi}_{0}$ with the GW190814\textendash like signal but at the $\mathrm{SNR}=200$.  The GR value (marked with a cross sign) with the $(2,2)$-mode waveform \pSEOBNR\, lies outside the 90\% credible level indicating a deviation from GR. }
	\label{fig:dchi0_high_snr}
\end{figure}

\begin{figure*}
	\begin{center}
		\includegraphics[width=1\textwidth]{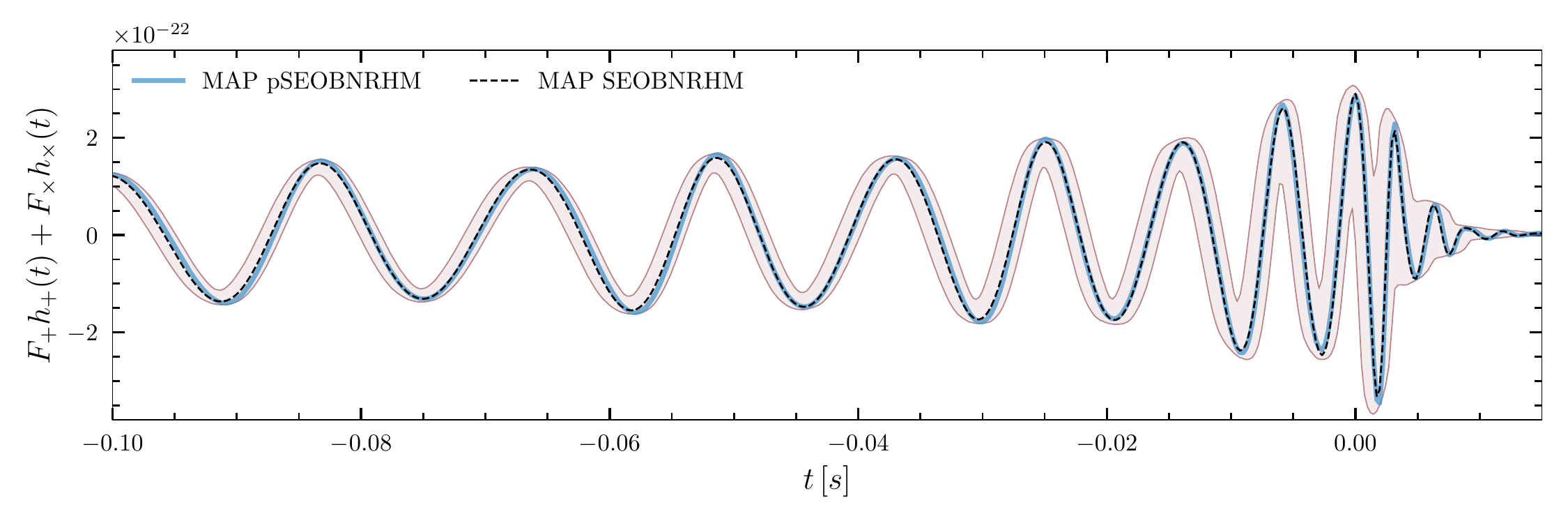}
	\end{center}
	\caption{The detector-strain outputs $F_+\,h_+(t) + F_\times\,h_\times(t)$ (where $F_+,F_\times$ are the antenna patterns at the LIGO Hanford detector) 
          of the \pSEOBNRHM{} (solid line) and GR \SEOBNRHM{} (dashed line) 
          MAP waveforms when a GW190412\textendash like signal is injected.  
          The \pSEOBNRHM{} waveform includes the deviation parameter $\delta \hat{\varphi}_{0}$. 
          The MAP \SEOBNRHM{} waveform is aligned with the  MAP \pSEOBNRHM{} waveform in a
          time interval corresponding to the orbital frequency $\{20, 30\}$ Hz. The
          \pSEOBNRHM{} and GR MAP waveforms are close to each other because their 
          MAP GR parameters are very similar. The shaded region denotes the 90\% uncertainty in the strain 
computed from the posterior samples of the parameters of \pSEOBNRHM{} in each time bin. }
	\label{fig:dchi_dphase}
\end{figure*}

We thus create two simulated signals, one with properties similar to
GW190412~\cite{GW190412} and another similar to GW190814~\cite{GW190814}, 
and choose our injection parameters to be the median values of these events~\cite{GW190814, GW190412} (see
Table~\ref{table:nedian_params})~\footnote{We set the spins to zero
  since their inferences were strongly prior-dependent.}. We choose 
the \pSEOBNRHM\, waveform model and a \emph{zero-noise}
configuration (i.e., we assume the data only contains the signal and no
noise, so as to remove noise-induced systematics and focus on
imperfections in waveform modeling). The properties of the two signals
are very different and we expect conclusions established through this
study to hold for a wide set of BBH signals. For the analysis of these
injections, we assume a three-detector network of the two LIGO and
Virgo detectors, with their respective sensitivities representative of
the third observing run (O3) configurations. Accordingly, we use the
PSDs provided in Ref.~\cite{KAGRA:2013rdx}. We set the minimum
frequency of the likelihood computation as $f_{\rm{low}}=20$ Hz. The
simulated signals are analyzed with two separate waveform models, \pSEOBNRHM\, 
and \pSEOBNR\,, to understand the effect of the
presence/absence of HMs in our waveform model on the results. We again
highlight here that we do not allow more than one deviation parameter
to vary during the inference analysis at any given time.

Before investigating these synthetic signals, we highlight a few
intriguing points about the black curves in
Fig.~\ref{fig:violin_GW190814}. First, for both the events, the
deviation parameter $\delta \hat{\varphi}_{-2}$ contains the GR value
(i.e., zero) at the tail of its posterior distributions. Note that $\delta \hat{\varphi}_{0}$
shows a similar albeit not as pronounced a behavior as $\delta
\hat{\varphi}_{-2}$. Additionally, for GW190814, the posterior distribution of
$\delta \hat{\varphi}_{5l}$ shows a small secondary peak at $\sim 0.62$, 
which gets enhanced upon the reweighing. These features {could be due to the particular noise realization even when using the Gaussian approximation}, or might be due to unaccounted
systematics, either in our understanding of noise properties around
the event, or our waveform modeling, especially in the high-mass
regimes. Thus, it would be worthwhile to have the results also
  from the corresponding simulated signals, so that some of these
  questions can be answered. We now discuss this. 

  In Fig.~\ref{fig:violin_GW190814} we show the results from the
  simulated signals next to the black curves (i.e., those from the
  actual events). We can see that the posterior distributions of $\delta
  \hat{\varphi}_{-2}$ obtained from the simulated signals with the HM
  waveform \pSEOBNRHM\, peak at zero.  Also, unlike in the case of the real
  event GW190814, $\delta \hat{\varphi}_{5l}$ from the simulated
  signal has a unimodal posterior distribution. These results suggest that the
  large bias seen in the case of real events are likely induced by {the
  noise content}~\footnote{The bias here is
    defined as the difference between the peak of the posterior distribution and
    the injected (true) value.}. Figure~\ref{fig:pn_bounds} shows 
   the 90\% bounds on the posterior distributions of
  the deviation parameters for both signals. The bounds from the
    real events and the corresponding simulated signals are different,
    however, they vary by much less than an order of magnitude. 
    This is interesting because there
    could be noise artifacts in the real data and because the
    simulated signals may not exactly represent the real events, yet
    we see that the bounds are of the same order.

We also show, in the same figures, the results obtained with the \pSEOBNR\, waveforms, 
which do not include HMs. One can see that the posterior distributions obtained with this waveform peak away
      from the injected value (zero) relative to the HM waveform \pSEOBNRHM{}. Among all, 
  the lower-order deviation parameters ($\delta \hat{\varphi}_{-2}$, $\delta \hat{\varphi}_{0}$,
      $\delta \hat{\varphi}_{1}$, $\delta \hat{\varphi}_{2}$, $\delta
      \hat{\varphi}_{4}$) have noticeable biases.  This suggests that
      neglecting HMs would compromise on accuracy, which would be
      consequential at high SNRs. In Fig.~\ref{fig:dchi0_high_snr}, we
      demonstrate this by repeating the analysis for the
        GW190814\textendash like signal but at SNR=200. As we can
      see, the GR prediction now lies outside the 90\% credible
      intervals when the HMs are not included in the recovery
      waveform.

      In addition to the accuracy, the inclusion of HMs also improves
      precision (i.e., the bounds) of the measurement of the deviation
      parameters (Fig.~\ref{fig:pn_bounds}). The improvement is
      marginal, but it is expected to improve significantly at high
      SNRs. Furthermore, consistent with our expectation, the lower PN-order 
      deviation parameters ($\delta \hat{\varphi}_{-2}$, $\delta
      \hat{\varphi}_{0}$, $ \delta \hat{\varphi}_{1}$, $\delta
      \hat{\varphi}_{2}$, $\delta \hat{\varphi}_{3}$, $\delta
      \hat{\varphi}_{4}$, $\delta \hat{\varphi}_{5l}$) are measured
      relatively better with the GW190814\textendash like signals as
      this system has lower total mass and thus more of the
      low-frequency inspiral falls within the bandwidth of the
      detectors. To give an example, the 90\% bound for $\delta
      \hat{\varphi}_{-2}$ obtained from GW190814\textendash like
      signals is $\sim \mathcal{O} (10^{-3})$, while from
      GW190412\textendash like signals it is $\sim \mathcal{O}
      (10^{-2})$, and thus there is an order of magnitude difference.
      On the other hand, for GW190412\textendash like signals which
      have higher total mass, the high PN deviation parameters
      ($\delta \hat{\varphi}_{6}$ and $\delta \hat{\varphi}_{7}$) are
      better measured.

      We end this section with a plot
        (Fig.~\ref{fig:dchi_dphase}) that show a typical \FTI{} waveform.  The shaded
        region shows the 90\% uncertainty in the \FTI{} waveforms  {from the analysis of a representative deviation
        parameter, in this case $\delta \hat{\varphi}_{0}$, with GW190412-like 
        injected signal. We first computed the waveforms corresponding to the posterior samples and then determine their 90\% uncertainty in each time bin.} We also plot the corresponding GR waveform which
        corresponds to the maximum \textit{a posteriori} (MAP)
        parameters in the GR analysis of the GW190412\textendash like
        signal, to explore the uncertainties of the \FTI{} waveforms
        around the \emph{true} GR waveform. We align the GR waveform
        with the \FTI {} MAP waveform at low frequency following the steps outlined in
        the Sec.~III A of Ref.~\cite{Ypan2011} (see, e.g., Eq.~(34) therein) in a time
        interval corresponding to the orbital frequency
        range $\{20, 30\}$ Hz~\footnote{The exact range of the orbital
          frequency chosen here for the alignment does not matter much
          since the parameters of the GR and non-GR waveforms are very close.}.
        Figure~\ref{fig:dchi_dphase} shows that the MAP \FTI{} and GR 
        waveforms are quite close to each other. This is because their GR parameters
        are very similar, as we will discuss below. 

\subsection{Optimization and systematics of the \FTI{} approach}
\label{robustness}

\begin{figure*}
	\begin{center}
		\includegraphics[width=0.49\textwidth]{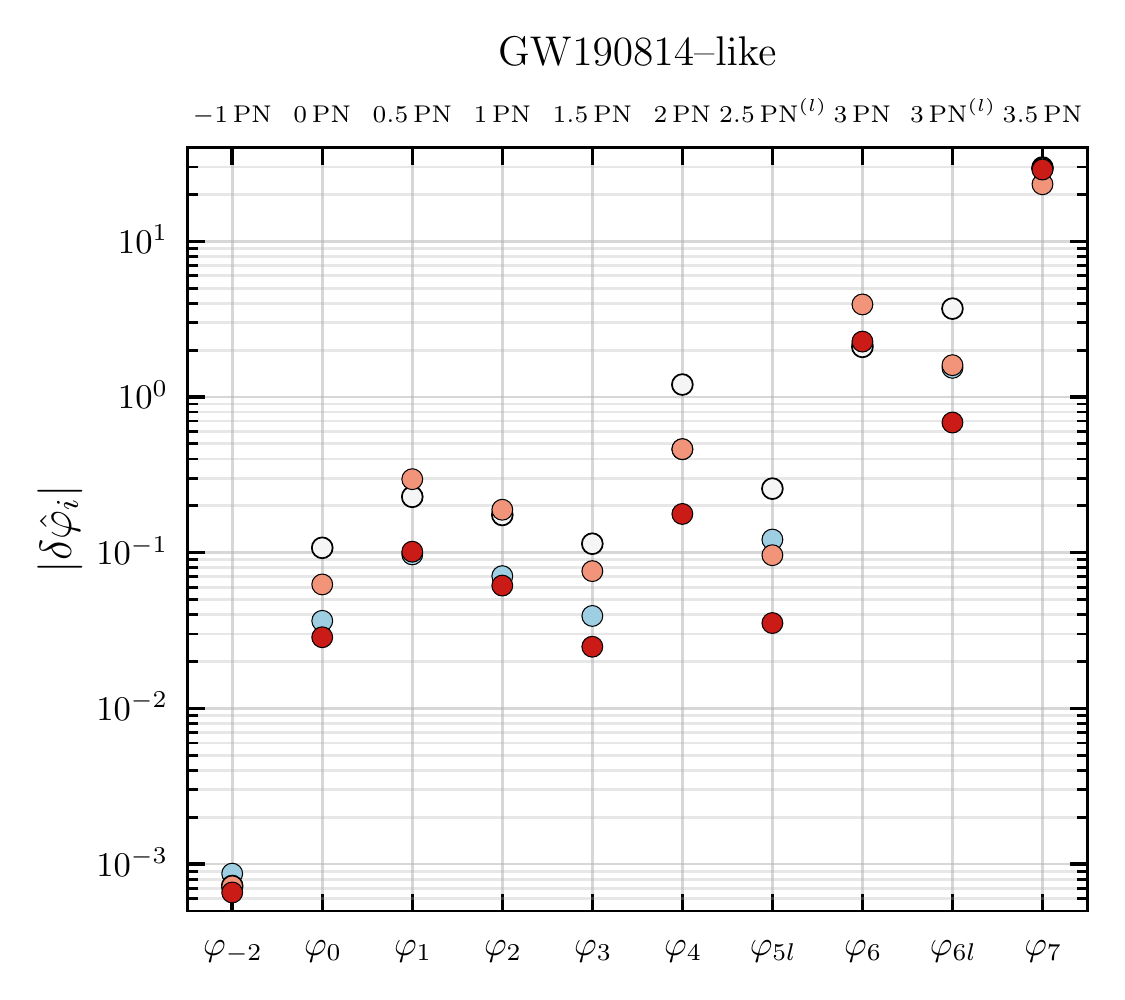}
                \includegraphics[width=0.49\textwidth]{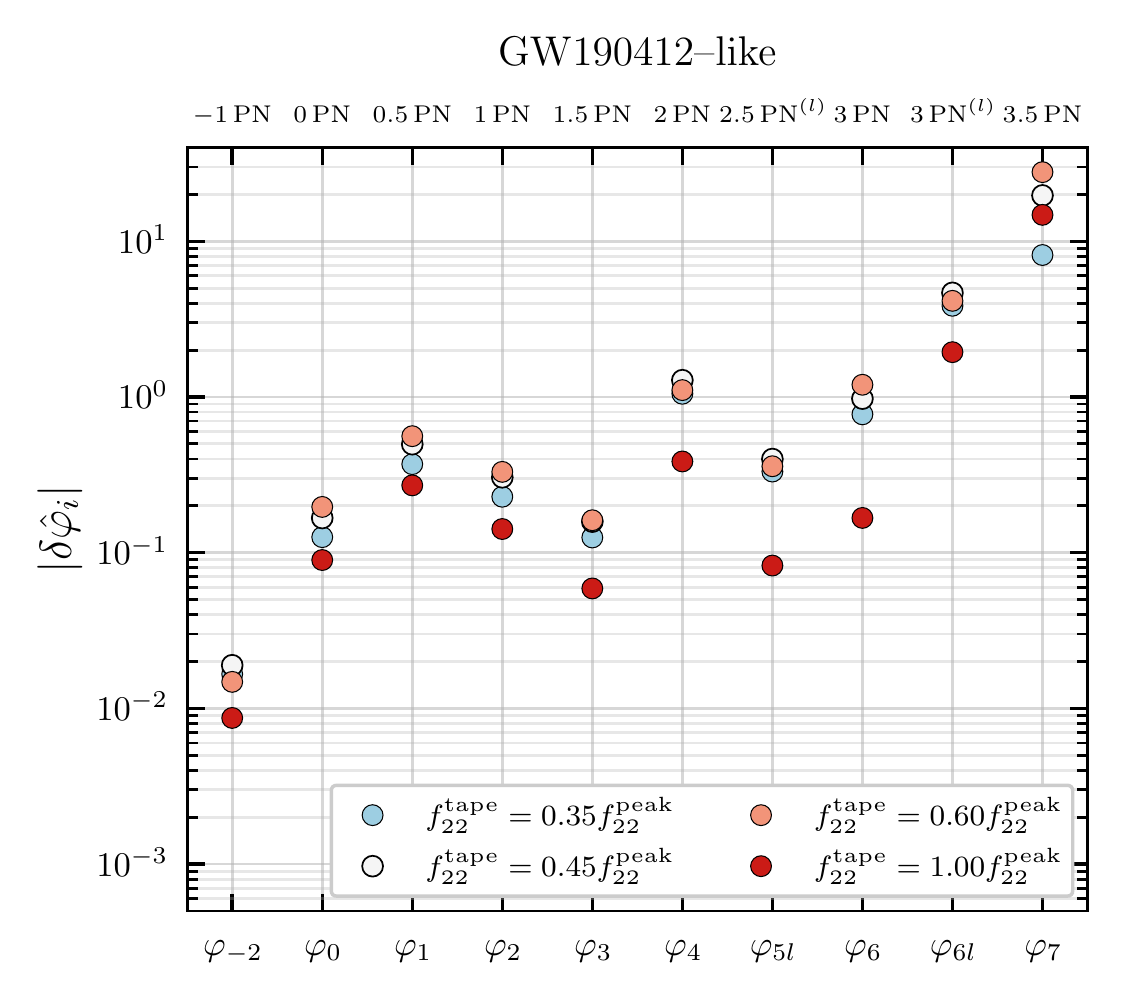}
	\end{center}
	\caption{The 90\% bounds on the deviation parameters from GW190814\textendash like (left panel) and GW190412\textendash like (right panel) signals when the tapering frequency $f_{22}^{\rm{tape}}$ is varied.}
	\label{fig:pn_bounds_f_win}
\end{figure*}

\begin{figure*}[htb!]
	\begin{center}
			\includegraphics[width=0.49\textwidth]{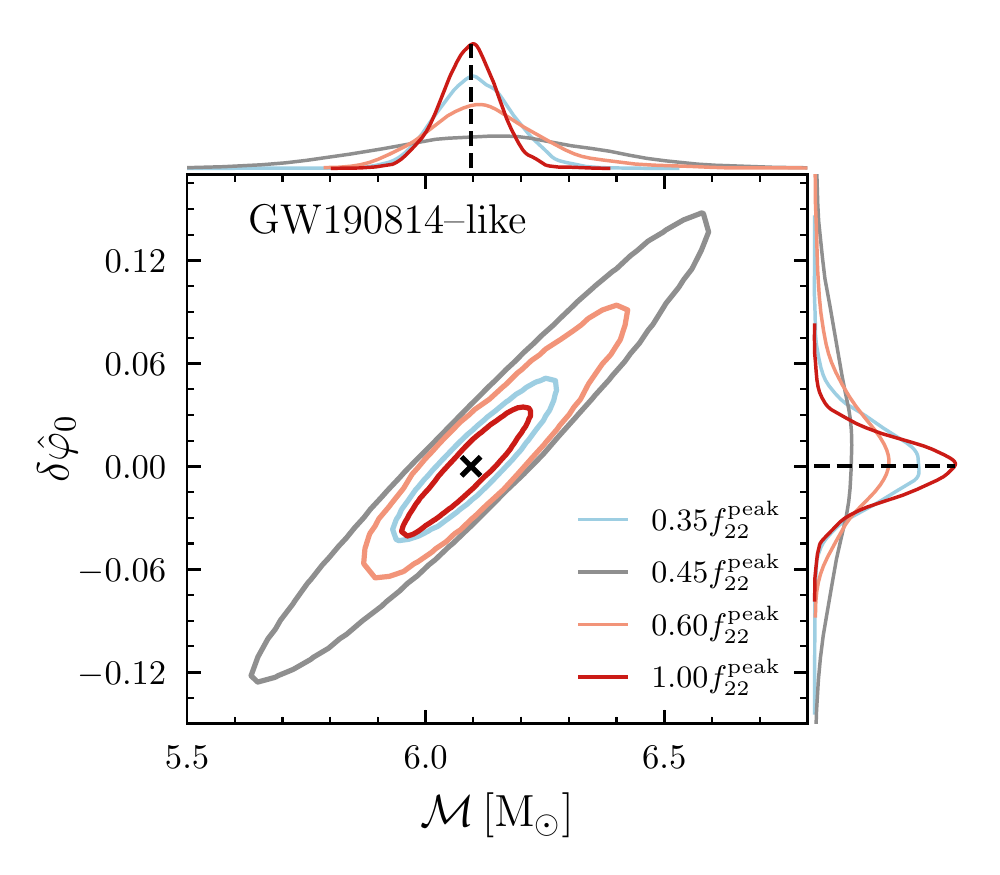}
			\includegraphics[width=0.49\textwidth]{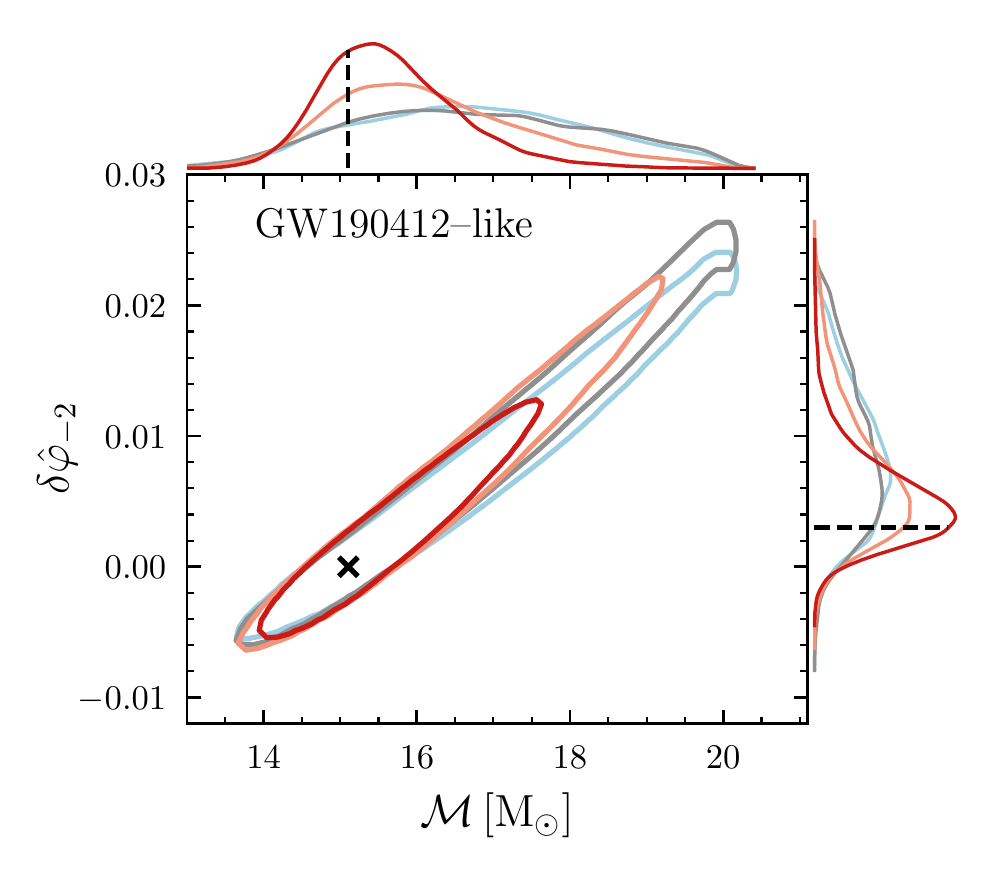}
	\end{center}
	\caption{The 2D joint posterior distribution between the deviation parameter $\delta \hat{\varphi}_{0}$ and the chirp mass $\mathcal{M}$ for the results presented in the left and right panels of Fig.~\ref{fig:pn_bounds_f_win}, respectively. The crosses and vertical dashed lines represent the injected values. The extent of the correlation between the chirp mass and the deviation parameters vary with the tapering frequency. }
	\label{fig:2d_post_f_win}
\end{figure*} 

\begin{figure*}[htb!]
	\begin{center}
		\includegraphics[width=0.49\textwidth]{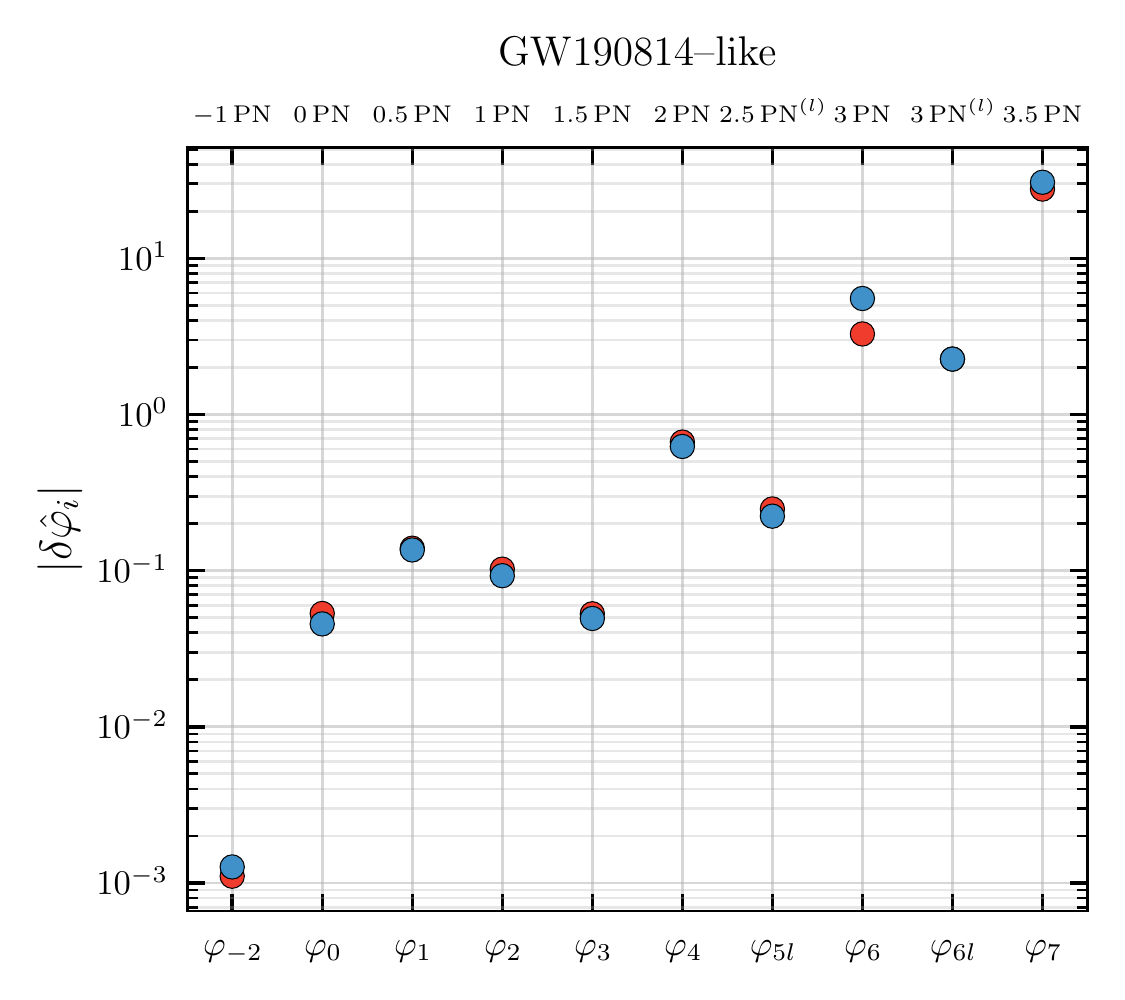}
		\includegraphics[width=0.49\textwidth]{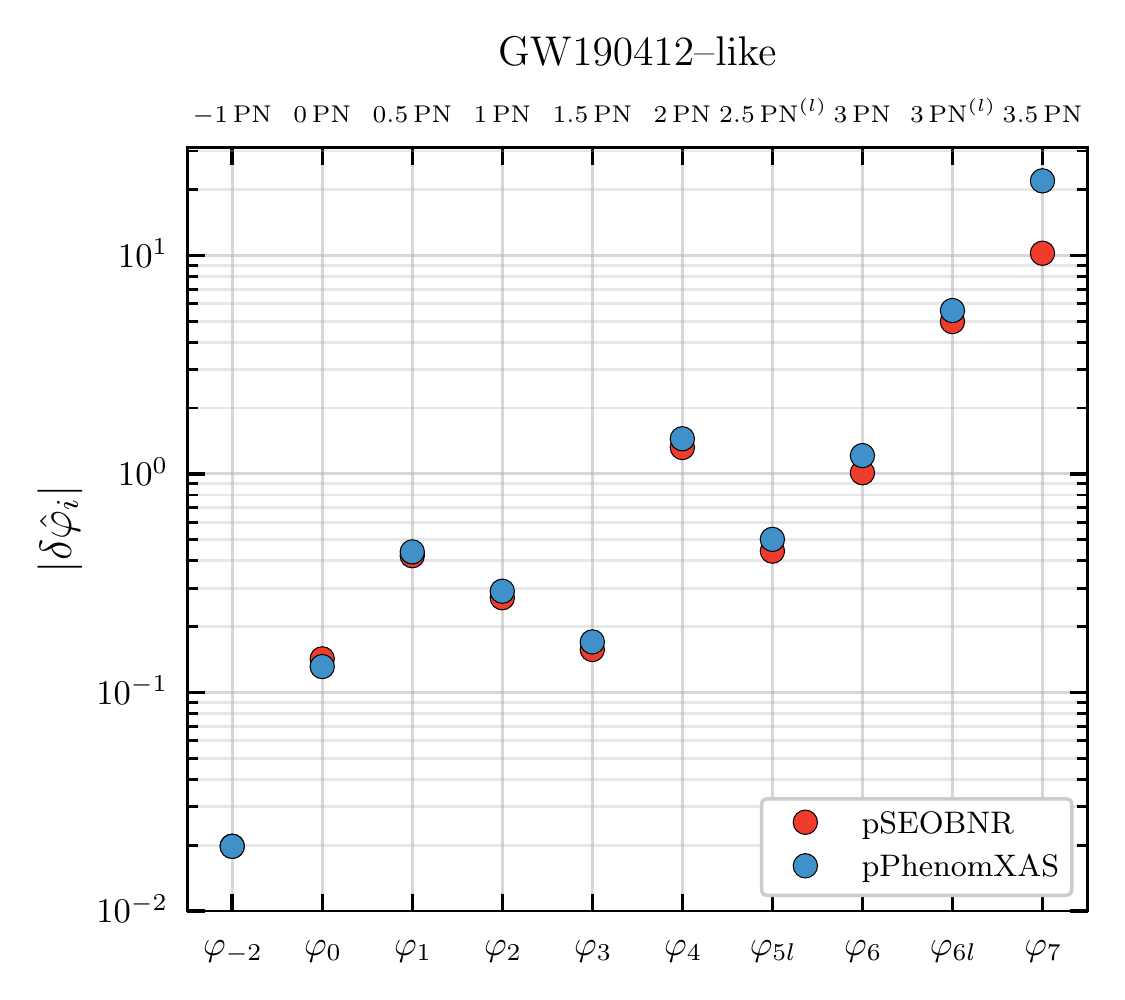}
	\end{center}
	\caption{The 90\% bounds on the deviation parameters from the simulated GW190814\textendash like  (left panel) and GW190412\textendash like (right panel) signals when two waveform families, namely, $\tt{pSEOBNR}$ and $\tt{pPhenomX}$, are used for the recovery.}
	\label{fig:pn_bounds_diff_waveform}
\end{figure*}

In this section we would like to investigate how to optimize the null tests enabled by the 
\FTI{} approach, and also explore possible systematics due to the settings, notably 
the choice of the tapering  frequency, and the family of waveform models adopted.

In Sec.~\ref{FTA:sec:Def} we have discussed a particular choice of the tapering frequency ($f^{\text{tape}}_{22} = 0.35
f_{22}^{\rm{peak}}$). Historically, this choice was motivated by comparisons of the \FTI{} results with \TIGER{} results~\footnote{{In the LIGO-Virgo analyses~\cite{TheLIGOScientific:2016src,TheLIGOScientific:2016pea,Abbott:2018lct,LIGOScientific:2019fpa,LIGOScientific:2020tif}, the \TIGER{} code~\cite{Agathos:2013upa} used the $\tt{IMRPhenomD}$ waveform model where the transition frequency between the inspiral and intermediate regions occurs close to  $f^{\text{tape}}_{22} = 0.35f_{22}^{\rm{peak}}$.}}. Nevertheless, given that the \FTI{}
method allows for flexibility in the tapering frequency, unlike
\TIGER, we want to understand the effects of varying it.  It is
also theoretically difficult to justify exactly where (i.e., at
  which orbital frequency) the inspiral ends. We thus allow the
tapering frequency to vary up until the (approximate) merger
frequency, $f_{22}^{\mathrm{peak}}$, and explore its impact on our
results. Varying the tapering frequency beyond the merger frequency
would not make much sense for an inspiral test. We note that \FTI{} analysis of 
BNS would not be affected by increasing the default tapering frequency, because for such
  systems, higher tapering frequency lie outside the sensitivity bandwidth of current 
GW detectors (e.g., $0.35 f_{22}^{\rm{peak}} \simeq 1350$ Hz for GW170817).


\begin{table}
	\centering
	\caption{{Fraction of the total SNR between $20$ Hz and the tapering frequency for the two simulated signals emplooyed here.}}
	\begin{tabular}{ccc} 
		{Tapering frequency} & GW190814\textendash like & GW190412\textendash like \\
\hline \rule{0pt}{1.1\normalbaselineskip}
		
	0.25$f_{22}^{\rm{peak}}$ & 83.8\% & 61.3\% \\[0.2cm]
		
	0.35$f_{22}^{\rm{peak}}$ & 92.1\% & 76.2\% \\[0.2cm]
		
           0.45$f_{22}^{\rm{peak}}$ & 95.4\% & 83.2\% \\[0.2cm]
		
		0.60$f_{22}^{\rm{peak}}$ & 97.4\% & 88.9\% \\[0.2cm]
		
		1.00$f_{22}^{\rm{peak}}$ & 99.3\% & 96.1\% \\[0.05cm]
		
		\end{tabular}
\label{Table:SNRs}
\end{table}

In Fig.~\ref{fig:pn_bounds_f_win} we show the change in the 90\%
  credible intervals of the deviation-parameter posterior distributions, as the
  tapering frequency is varied. More  specifically, the exact change in the bounds depends on the
  underlying signal itself, for example, for the high--total-mass
  GW190412\textendash like system (right panel), with less
  inspiral in the detectors' frequency bands, bounds on the lower
  PN order deviation parameters ($\delta \hat{\varphi}_{-2}$, 
$\delta \hat{\varphi}_{0}$, $\delta \hat{\varphi}_{1}$, $\delta
  \hat{\varphi}_{2}, \hat{\varphi}_{3}$) change at most by a factor of $\sim 2\mbox{-}3$ (if we push the 
tapering frequency up to the peak of $h_{22}$). This makes sense because
  those deviation parameters affect the waveform significantly only at low frequencies and thus
  increasing the tapering frequency does not make much difference to
  the results. The change on the bounds increases (up to $\sim 5$) for the higher PN deviation parameters, 
  and becomes the largest for $\delta \hat{\varphi}_{6}$, for which the variation is $\sim 7$. 
  Extending the tapering frequency up to close to merger increases the available SNR {(see Table \ref{Table:SNRs})}, and improves 
the measurement of the higher PN deviation parameters $\delta \hat{\varphi}_{4}$, $\delta \hat{\varphi}_{5 \ell}$, 
$\delta \hat{\varphi}_{6}$.

For low-total-mass GW190814\textendash like
  systems, as one can see from the left panel of
  Fig.~\ref{fig:pn_bounds_f_win}, for almost all deviation parameters,
  the tapering frequency has a somewhat larger impact on the bounds. Bounds
  for the lower PN order deviation parameters ($\delta
  \hat{\varphi}_{-2}$, $\delta \hat{\varphi}_{0}$, $\delta
  \hat{\varphi}_{1}$, $\delta \hat{\varphi}_{2}$, and $\delta
  \hat{\varphi}_{3}$) change by a factor of $1\mbox{-}5$, while the
  higher-order ones change by a factor as large as $\sim 7$ 
(for $\hat{\varphi}_{5 \ell}$). Also in this case, extending the tapering 
frequency up to merger, can improve the measurement of the high PN deviation 
parameters. The above studies suggest that to optimize the \FTI{} framework for BBHs, it would 
be beneficial to obtain a tapering frequency that depends on {the available SNR 
and the number of GW cycles up to the peak of the waveform. Eventually, those quantities 
depend on the total mass of the binary, mass ratio and spins.}

As stated above, the chirp mass $\mathcal{M}$ generally correlates 
  with the deviation parameters, a property discussed in more detail
  in the following subsection. This means that the measurement
  of the chirp mass would get affected by the variation of the
  tapering frequency. This can be seen in 
    Fig.~\ref{fig:2d_post_f_win}. The left panel shows the correlation
    of $\mathcal{M}$ with $\delta \hat{\varphi}_{0}$ for
    GW190814\textendash like systems, while the right panel shows the
    correlation with $\delta \hat{\varphi}_{-2}$ for
    GW190412\textendash like systems. From the left
    panel we observe that the variation in the tapering frequency only
    affects the width of the chirp-mass posterior distribution, while the right
    panel shows that, additionally, there could arise some biases (even
    in the deviation parameter $\delta \hat{\varphi}_{-2}$) if the
    tapering frequency is too low. The reason for this could be
    that, for relatively high-mass systems, the number of GW
      cycles in the detectors' frequency bands is already low to start
      with, and thus using too low tapering frequencies would
      essentially make the inference almost insensitive to $\delta
      \hat{\varphi}_{-2}$. We note that for the GW190412-like source, 
the bias is quite reduced when we use the tapering frequency at the peak 
of the (2,2) mode. This is because for this 
high-mass binary the last few cycles before merger can increase significantly 
the SNR accumulated. Hence, if one should find evidence for a
      violation of GR, one must first check for a possible bias of the
      \FTI{} results by varying the tapering frequency. We find that
    the other GR parameters, besides the chirp mass, remain mostly
    unaffected. This could be because they play a subdominant role in
    the orbital dynamics during the inspiral.

Finally, so far we have employed for our analyses the EOB-based
      waveform models: $\tt{SEOBNRHM}$ and $\tt{SEOBNR}$. We show 
      in Fig.~\ref{fig:pn_bounds_diff_waveform}
      the results obtained from the two simulated signals when recovered with the aligned-spin phenomenological 
      waveform \pPhenomX{}, in addition to the $\tt{SEOBNR}$
      waveform.  Both models only contain the $(2,2)$ mode. 
      As we can see, there is really no significant difference between the
      bounds obtained from these two different waveform families, except
      for $\delta \hat{\varphi}_{7}$ with the higher-mass
      GW190412\textendash like system (perhaps due to modeling
      differences at high frequencies). In addition, the details of
      the individual posterior distributions (not shown here) are very similar. We
    thus expect that the results established in this work do not 
     depend much on the underlying family of GR waveform models used, as long 
as they have comparable accuracy to numerical relativity.

\subsection{Impact of the \FTI{} approach on the GR parameters}

\begin{figure*}[htb!]
	\begin{center}
		\includegraphics[width=1\textwidth]{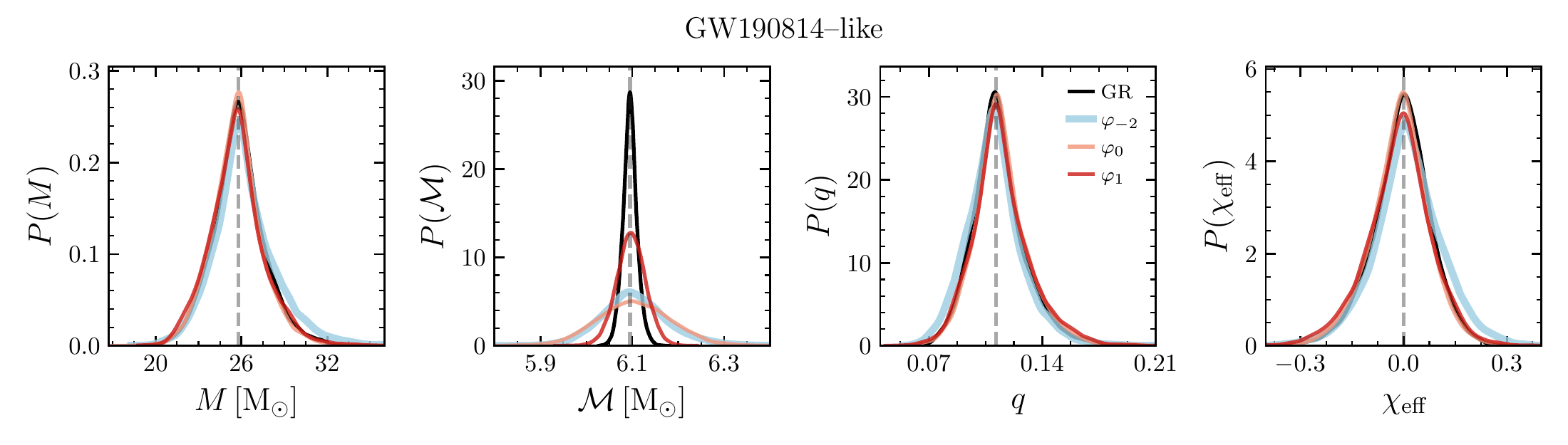}
		\includegraphics[width=1\textwidth]{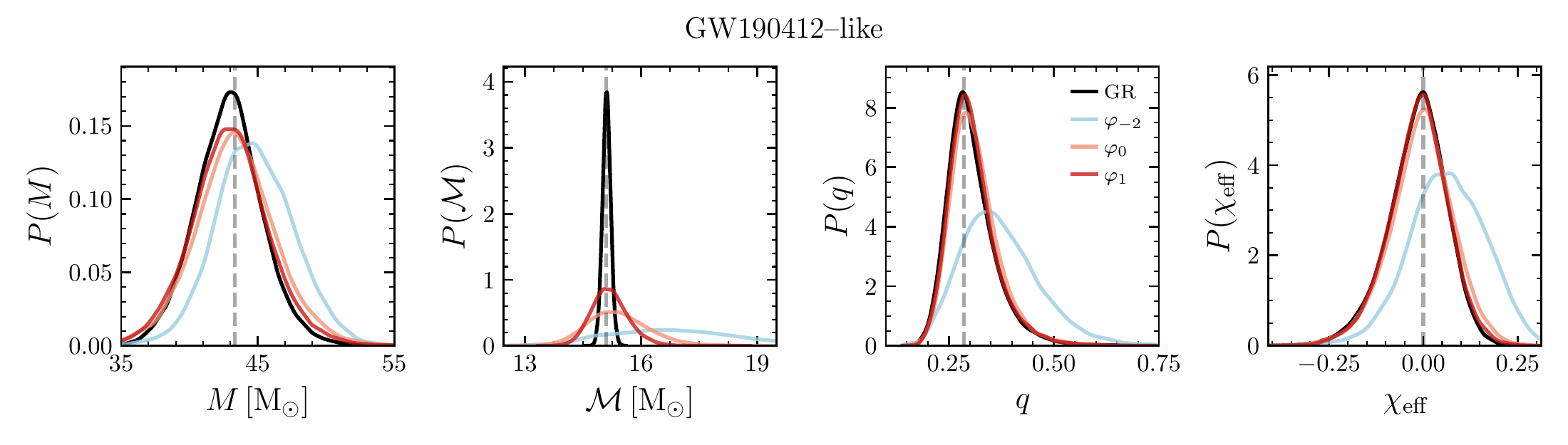}
	\end{center}
	\caption{Posterior distributions of several binary's parameters for the simulated GW190814\textendash like signal (upper panel) and the GW190412\textendash like signal (lower panel) when 
 the GR \SEOBNRHM{} model and the \pSEOBNRHM{} model with deviations parameters $\delta \hat{\varphi}_{-2}$, $\delta \hat{\varphi}_{0}$, and $\delta \hat{\varphi}_{1}$ are used to recover the signal. The vertical dashed lines represent the true value of the injections. The strong correlation between these deviation parameters and the chirp mass $\mathcal{M}$ leads to the broadening of the chirp mass posteriors and sometime causes a bias as well.}
	\label{fig:GR_params_1}
\end{figure*}

\begin{figure*}[htb!]
	\begin{center}
		\includegraphics[width=1\textwidth]{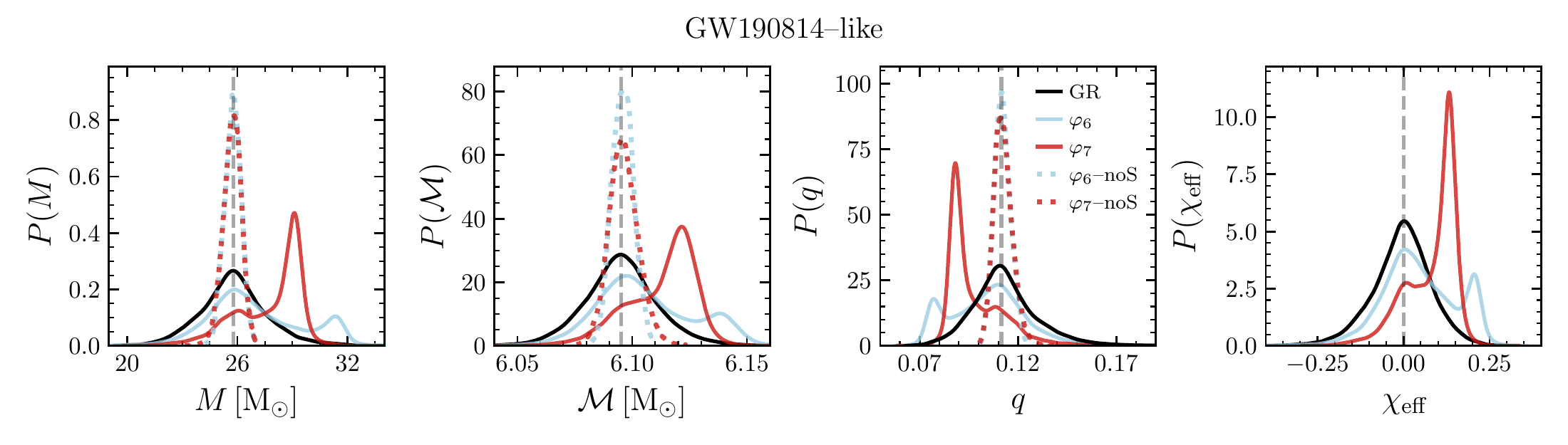}
		\includegraphics[width=1\textwidth]{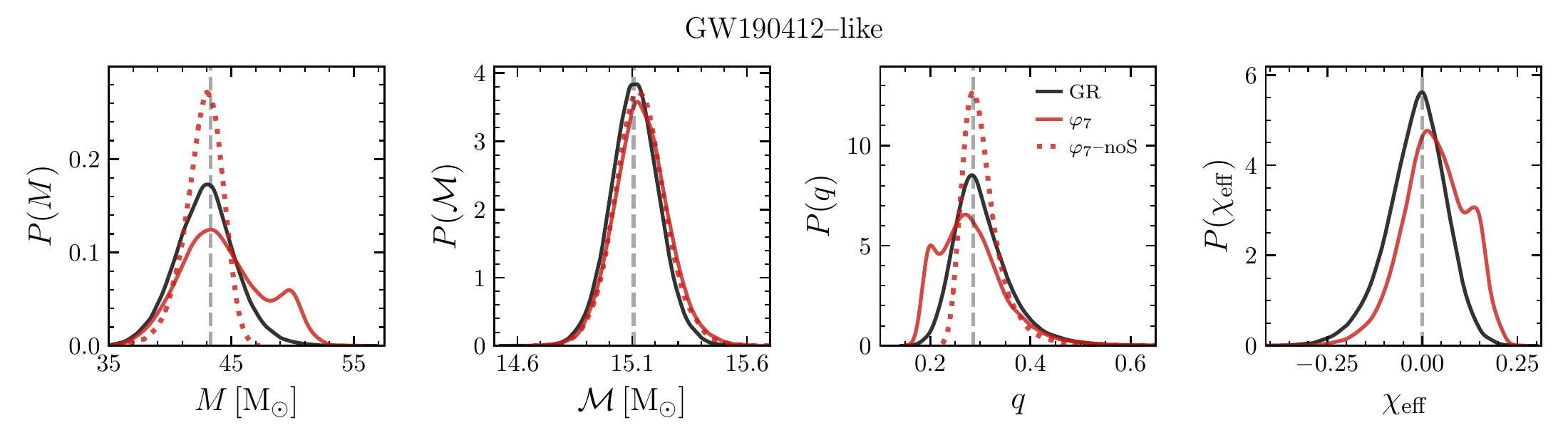}
	\end{center}
	\caption{Same as Fig.~\ref{fig:GR_params_1} but for the deviation parameters $\delta \hat{\varphi}_{6}$ and $\delta \hat{\varphi}_{7}$. The dotted lines represent the posteriors obtained with the non-spinning (noS) \pSEOBNRHM{} waveforms. The bimodalities in the GR parameters arise as a consequence of strong correlations between masses and spins induced by the deviation parameters $\delta \hat{\varphi}_{6}$ and $\delta \hat{\varphi}_{7}$.}
	\label{fig:GR_params_2}
\end{figure*}

In Figs.~\ref{fig:GR_params_1} and \ref{fig:GR_params_2} we show the posteriors of the 
GR parameters of the two simulated signals introduced above, when recovering them with 
the GR \SEOBNRHM{} model and the \pSEOBNRHM{} model. For the cases involving the parameters 
$\delta \hat{\varphi}_{-2}$ and $\delta \hat{\varphi}_{0}$, we observe that the measurement 
uncertainty of the chirp mass increases significantly, while other GR parameters, e.g., 
the total mass ($M$), mass ratio ($q$), or effective spin ($\chi_{\rm{eff}}$), remain unaffected. 
For the GW190412\textendash like signal, the addition of $\delta \hat{\varphi}_{-2}$ introduces 
a bias in the measurement of the GR parameters. These biases, however, reduce 
if a tapering frequency higher than the current default of $0.35 f_{22}^{\mathrm{peak}}$ is used 
for the analysis, as demonstrated in the previous subsection. Hence, the biases 
appear to be the consequence of an insufficient number of GW cycles or SNR.

The fact that the chirp-mass measurement is heavily correlated or degenerate with the measurement of some non-GR parameters can be understood by looking at the \FTI{} formulation itself. Restricting ourselves to the $(2,2)$-mode, using the definition of $v$ from Eq.~\eqref{eq:orb_freq}, 
the $n/2$-PN-order coefficient in Eq.~\eqref{eq:deltaPN} that contributes to the total phase correction is

\begin{equation}
\psi_{n}(f;\bm{\theta}) = \dfrac{3}{128 (\mathcal{M} \pi f)^{5/3}}\,  \psi_n^{(\text{GR})} (\bm{\theta}) (1+\delta \hat{\varphi}_{n}),
\end{equation}
where $\delta \hat{\varphi}_{n}$ is the $n/2$-PN-order deviation parameter. For the leading-order term $n=0$, this implies
\begin{equation}
\psi_{0}(f; \bm{\theta}) = \dfrac{3}{128 (\mathcal{M} \pi f)^{5/3}} \,(1+\delta \hat{\varphi}_{0}) \label{psi0}
\end{equation}
since $\psi_0^{(\text{GR})} =1$. For the -1PN and 0.5PN phase corrections (i.e., $n=-2, 1$), which are absolute corrections since they are 
absent in GR, the expressions read
\begin{equation}
\psi_{-2}(f; \bm{\theta}) = \dfrac{3}{128 (\mathcal{M} \pi f)^{5/3}} \, \delta \hat{\varphi}_{-2} \label{psi-2} ,
\end{equation}
and
\begin{equation}
\psi_{1}(f; \bm{\theta}) = \dfrac{3}{128 (\mathcal{M} \pi f)^{5/3}} \, \delta \hat{\varphi}_{1} . \label{psi1}
\end{equation}
It becomes clear from Eqs.~\eqref{psi0}, \eqref{psi-2}, and \eqref{psi1} that the deviation parameters $\delta \hat{\varphi}_{-2}$, $\delta \hat{\varphi}_{0}$ and $\delta \hat{\varphi}_{1}$ are degenerate with the chirp mass, as Fig.~\ref{fig:2d_post_f_win} shows. There are also correlations between the chirp mass (and other binary parameters) and the deviation parameters at higher PN orders, but they are milder. Indeed the addition of $\delta \hat{\varphi}_{2}$, $\delta \hat{\varphi}_{3}$, $\delta \hat{\varphi}_{4}$, $\delta \hat{\varphi}_{5l}$ and $\delta \hat{\varphi}_{6l}$ do not affect estimates of GR parameters in any noticeable fashion, as we have verified using the results in Figs.~\ref{fig:GR_params_1} and \ref{fig:GR_params_2}.

However, for the highest PN-order deviation parameters, $\delta \hat{\varphi}_{6}$ and $\delta \hat{\varphi}_{7}$, posterior distributions of GR parameters can show features like bimodalities, depending on the underlying signal, see Fig.~\ref{fig:GR_params_2}. This is because, unlike the cases of $n=-2,0,1$, for values of $n \geq 2$ the PN coefficients $\psi_n^{(\text{GR})}$ also depend on the intrinsic properties, in particular the symmetric mass ratio and the spins. In fact, the bimodalities observed in the $\delta \hat{\varphi}_{6}$ and $\delta \hat{\varphi}_{7}$ cases disappear when we perform the analysis with non-spinning waveforms, shown by the dotted lines in Fig.~\ref{fig:GR_params_2}. This suggests that the deviation parameters $\delta \hat{\varphi}_{6}$ and $\delta \hat{\varphi}_{7}$ induce strong correlations between the GR parameters when the binary is spinning. We notice that the amount of bimodality also depends on the tapering frequency, notably on the GW cycles and SNR.

\section{Application of the \FTI{} approach to a binary neutron star}
\label{FTA:sec:BransDicke}

Here we consider the application of the \FTI{} approach to a specific
alternative theory of GR: the Jordan--Fierz--Brans--Dicke (JFBD)
scalar--tensor theory~\cite{Jordan:1955,Fierz:1956zz,Brans:1961sx}. Initially
formulated in the mid-20th century, JFBD gravity was the very first
scalar-tensor theory---a theory in which gravity is mediated by both a
tensor (the metric) and a scalar. Since then, significant work has
been done to extend this notion beyond JFBD theory to broader, more
generic classes of scalar-tensor theories (e.g., Horndeski
theories~\cite{Horndeski:1974wa}, Beyond Horndeski
theories~\cite{Gleyzes:2014dya}, Degenerate Higher-Order Scalar-Tensor
theories~\cite{Langlois:2015cwa,Langlois:2017mdk}). Yet, despite its
simplicity, JFBD gravity remains relevant today, though more as a
pedagogical archetype of modified gravity than as a truly viable
alternative to GR.  In this vein, constraining JFBD theory with a
particular experiment offers an easily understood benchmark of its
sensitivity to deviations from GR.

The action for JFBD  gravity written in the Jordan frame is given by
\begin{align}
S = \int d^4 x \frac{\sqrt{-\tilde{g}}}{16 \pi}\left( \phi \tilde{R} - \frac{\omega_\text{BD}}{\phi} \tilde{g}^{\mu \nu}\partial_\mu \phi \partial_\nu \phi \right)+ S_m[\tilde{g}_{\mu \nu}, \psi],
\end{align}
where $\phi$ is a massless scalar field, $\omega_\text{BD}$ is a dimensionless coupling constant\footnote{JFBD is also commonly known as simply Brans-Dicke gravity (BD); following the standard convention in the literature, we adopt this abbreviation when denoting the coupling constant $\omega_\text{BD}$.}, and $S_m$ represents the action for matter fields $\psi$ minimally coupled to the metric $\tilde{g}_{\mu \nu}$. (Here $\psi$ should not be confused with the GW modes $\psi_{\ell m}$ introduced earlier.)
Alternatively, the action can be rewritten in the Einstein frame by performing the conformal transformation $g_{\mu \nu} \equiv \phi \tilde{g}_{\mu \nu}$
\begin{align}
S = \int d^4 x \frac{\sqrt{-g}}{16 \pi}\left( R - 2 g^{\mu \nu}\partial_\mu \varphi \partial_\nu \varphi \right)+ S_m[ e^{-2\alpha_0 \varphi} g_{\mu \nu}, \psi],\label{FTA:eq:EinsteinFrame}
\end{align}
where we have defined the dimensionless parameter $\alpha_0\equiv (3+ 2 \omega_\text{BD})^{-1/2}$ and introduced the scalar field $\varphi \equiv   \log(\phi)/(2 \alpha_0)$; note that $\alpha_0$ is non-negative and that we have implicitly assumed that $\omega_\text{BD}>-3/2$. (Note that $\varphi$ is unrelated to the \FTI{} deviation parameters.) In the limit that $\alpha_0 \rightarrow 0$ ($\omega_\text{BD} \rightarrow \infty$), the scalar field decouples from the metric and matter, and JFBD theory reduces to GR with an additional massless scalar that is minimally coupled to gravity only. 
{The most accurate constraints on this parameter come from the Doppler tracking of the Cassini spacecraft through the Solar System~\cite{Bertotti:2003rm} 
$\alpha_0 <4 \times 10^{-3}$ ($\omega_\text{BD} > 4\times 10^4 $), and binary pulsar obervations~\cite{Damour:1992we,Damour:1993hw,Wex:2014nva,Anderson:2019eay}. In particular, the most recent results obtained in Ref.~\cite{Kramer:2021jcw} 
with 16 years of observation of the double-pulsar J0737-3039 yield the bounds $\alpha_0 = 0.004083$, and $= 0.003148$ at $95\%$ credible level, when the stiff MPA1 and soft WFF1 EOS are employed, respectively.}

The recent advent of GW astronomy offers a new avenue to test gravity in the relativistic regime. The majority of GWs observed by LIGO and Virgo thus far were generated by the coalescence of  BBHs; several tests of GR have already been conducted using these observations~\cite{TheLIGOScientific:2016src, TheLIGOScientific:2016pea, Yunes:2016jcc, LIGOScientific:2019fpa}. However, Hawking famously showed that stationary BHs in JFBD theory must have a trivial scalar profile, and thus are indistinguishable from the analogous solutions in GR~\cite{Hawking:1972qk}. Although there are some possible scenarios that evade this no-hair theorem (see, e.g., Ref.~\cite{Sotiriou:2015pka}), binary systems composed of BHs are generally expected to behave identically in JFBD theory and GR, and thus GWs from such systems are unable to constrain this scalar-tensor theory.

Unlike BHs, NSs source a nontrivial scalar field in JFBD gravity, and thus BNS systems can be used to constrain $\alpha_0$.  In this section, we use the first GW observation of a coalescing BNS---GW170817~\cite{TheLIGOScientific:2017qsa}---to constrain $\alpha_0 $ at the 68\% (and 90\%) credible levels. Though the constraint from GW170817 is not as strong as those previously quoted from other experiments, this result represents the first bound directly from the highly dynamical (orbital velocities $v \sim 10^{-1}$) and strong-field (Newtonian potential $\Phi_\text{Newt} =M/R \sim 10^{-1}$) regime of gravity.

This section is organized as follows. In Sec.~\ref{sec:Signature}, we detail the GW signature of JFBD theory in BNSs. Then, in Sec.~\ref{FTA:sec:BDconstraints}, we present two Bayesian analyses to constrain $\alpha_0$ with GW170817: the first directly uses the theory-agnostic analyses presented 
in Sec.~\ref{FTA:sec:Def}, while the second is tailored specifically to test JFBD gravity.

\subsection{Gravitational-wave signature of JFBD gravity}\label{sec:Signature}

The predominant differences in GWs produced in JFBD gravity as compared to GR
stem from the fact that only the latter respects the strong
  equivalence principle.  This principle extends the universality of
free fall by test particles implied by the Einstein equivalence
principle to also include self-gravitating bodies; unlike in GR, the motion 
of a body through spacetime depends on its internal gravitational interactions 
(i.e., its composition) in scalar-tensor theories like JFBD gravity.  This section details how this 
violation of the strong equivalence principle impacts the GWs produced by binary
systems in JFBD gravity.  This alternative theory of gravity falls within the
class of scalar-tensor theories for which PN predictions have been computed. 
Though those results are available at next-to-next-to-leading PN order (and even higher-order PN calculations have been made recently~\cite{Sennett:2016klh,Bernard:2018hta,Bernard:2018ivi,Bernard:2022noq}), we will assume that
$\alpha_0$ is sufficiently small that we can neglect all but the
leading-order PN effects when describing the signature of JFBD gravity in a
gravitational waveform.

\begin{table*}\caption{One-dimensional polynomial fits of the normalized NS coupling $\alpha_i/\alpha_0$ as a function of NS mass $m_i$ (in units of $M_\odot$) for various EOSs. The relative error amounts to less than 5\% over the area $(m_i,\alpha_0) \in [0.5\,M_\odot,2.0\,M_\odot] \times [0.001,1.0]$.}\label{FTA:tab:1dfit}
\begin{center}
\begin{tabular}{cr@{~~~~}r@{}r@{}r@{}r@{}r}
\multicolumn{2}{c}{EOS~~~~} & \multicolumn{5}{c}{fit for $[\alpha_i/\alpha_0](m_i)$} \\
\hline \\ [-2.5ex]
sly & \cite{Douchin:2001sv} & $- 0.726798$ & $- 0.749029 \, m_i$ & $+ 1.270944 \, m_i^2$ & $- 0.728710 \, m_i^3$ & $+ 0.161002 \, m_i^4$ \\
eng & \cite{Engvik:1995gn} & $- 0.817884$ & $- 0.393375 \, m_i$ & $+ 0.772615 \, m_i^2$ & $- 0.435306 \, m_i^3$ & $+  0.095059 \, m_i^4$ \\
H4 & \cite{Lackey:2005tk} & $- 0.613880$ & $- 1.210074 \, m_i$ & $ + 1.836631 \, m_i^2$ & $- 1.056595 \, m_i^3$ & $+ 0.228102 \, m_i^4$ \\
\end{tabular}
\end{center}
\end{table*}

\begin{table*}\caption{Two-dimensional polynomial fits of the normalized NS coupling $\alpha_i / \alpha_0$ as functions of the Brans-Dicke parameter $\alpha_0$ and NS mass $m_i$ (in units of $M_\odot$)  for various EOSs. The relative error amounts to less than 1\% over the area $(m_i,\alpha_0) \in [0.5\,M_\odot,2.0\,M_\odot] \times [0.001,1.0]$.}\label{FTA:tab:2dfit}
\begin{center}
  \begin{tabular}{cr@{~~~~}r@{}r@{}r@{}r@{}r@{}r}
\multicolumn{2}{c}{EOS~~~~} & \multicolumn{6}{c}{fit for $[\alpha_i / \alpha_0](m_i, \alpha_0)$} \\
\hline \\ [-2.5ex]
sly & \cite{Douchin:2001sv} & $- 0.92569$ && $+ 0.22258 \, \alpha_0 m_i $ & $+ 0.13329 \, m_i^2 $ & $- 0.15151 \, \alpha_0 m_i^2$ & \\
eng & \cite{Engvik:1995gn} & $- 0.97423$ & $+ 0.15584 \, m_i$ & $+ 0.18527 \, \alpha_0 m_i $ && $- 0.11739 \, \alpha_0 m_i^2 $ & $+ 0.024333 \, m_i^3$ \\
H4 & \cite{Lackey:2005tk} & $- 0.93341$ && $+ 0.19073 \, \alpha_0 m_i$ & $+ 0.10270 \, m_i^2 $ & $- 0.11284 \, \alpha_0 m_i^2$ &
\end{tabular}
\end{center}
\end{table*}

The dominant effect on the inspiral from the new scalar introduced in JFBD gravity is the emission of dipole radiation, which enters into the phase evolution at -1PN order.
In the notation of the \FTI{} framework and using the fact that the quadrupolar GR radiation dominates over the dipolar one for small $\alpha_0$~\cite{Sennett:2016klh}, this contribution is given by
\begin{align}
\delta \hat{\varphi}_{-2} = -\frac{5(\alpha_1 - \alpha_2)^2}{168} +\Ord\left(\alpha_0^4\right),\label{FTA:eq:DipoleTerm}
\end{align}
where $\alpha_i$ is the scalar charge of body $i$, defined as
\begin{align}
  \alpha_i \equiv - \frac{d \log m_i(\varphi)}{d \varphi},
\end{align}
where $m_i(\varphi)$ is the gravitational mass of body $i$ measured in the Einstein frame ($i=1,2$). That a body's mass depends on the local value of the scalar field is unsurprising given the form of Eq.~\eqref{FTA:eq:EinsteinFrame}; a shift in $\varphi$ modulates the physical metric $e^{- 2 \alpha_0 \varphi} g_{\mu \nu}$ that effects gravity upon the matter fields $\psi$, and thus also modulates any body's gravitational mass.
This dependence is an explicit manifestation of violation of the strong equivalence principle.
Note that in the limit that a body has no self-gravity (i.e., the test-body limit), the functional form of $m_i(\varphi)$ simplifies significantly to
\begin{align}
m_i^{\text{(test body)}}(\varphi) = e^{- \alpha_0 \varphi} m_i^{\text{(test body)}}(\varphi=0) ,
\end{align}
and thus its scalar charge reduces to
\begin{align}
  \alpha_i^{\text{(test body)}} = \alpha_0.
\end{align}

As strongly self-gravitating bodies, violations of the strong equivalence principle are particularly pronounced in NSs.
This violation manifests as a scalar charge that differs significantly from the test-body charge $\alpha_0$.
As the scalar charges of a binary's constituents---or rather their difference, the scalar dipole---control the dominant effect on the GW signal in JFBD gravity, we devote the remainder of this subsection to computing these quantities for various NSs.

We consider spherically symmetric, static solutions sourced by a perfect fluid as a model for an isolated, nonspinning NS.
Under these assumptions, the field equations for Eq.~\eqref{FTA:eq:EinsteinFrame} reduce to the Tolman-Oppenheimer-Volkoff (TOV) equations, given in the Einstein frame in Ref.~\cite{Damour:1993hw}.
These solutions are parameterized by three degrees of freedom; for our purposes, these are most clearly manifested as the (i) background scalar field $\varphi_0$, that is the scalar field far from the NS, (ii) the NS EOS, and (iii) the NS mass.\footnote{We define the NS mass as the tensor mass $m_T$ introduced in Ref.~\cite{Lee:1974pt} because---as shown in that reference---it obeys the same conservation laws as the ADM mass in GR.}
Note that the asymptotic scalar field $\varphi_0$ can be set to zero without loss of generality by rescaling the Jordan-frame bare gravitational constant $\tilde G$ accordingly, that is $\varphi_0 \rightarrow 0 \Rightarrow \tilde{G}\rightarrow \tilde{G}e^{2\alpha_0 \varphi_0}$.
The remaining degrees of freedom can be mapped to boundary conditions for the matter and scalar field at the origin with a numerical shooting method~\cite{Press:1992:NRC:148286}.
These conditions are parameterized by the central pressure $P_c$ and the scalar field $\varphi_c$, which serve as the inputs for numerically integrating the TOV equations. (Note that in this section $\varphi_c$ is different from the parameter used in Sec.~\ref{FTA:sec:Def} 
to define the binary's orientation.) The details for extracting the mass and scalar charge from the numerical solutions of the NS interior are given in
Ref.~\cite{Damour:1993hw}.

Ultimately, we would like to combine the numerical calculations of the NS scalar charge outlined above with their anticipated effect on the GW signal~\eqref{FTA:eq:DipoleTerm} to constrain JFBD theory.
However, evaluating the scalar charge directly for every point visited by the stochastic sampling algorithms used for parameter estimation would require an unreasonable amount of computational resources.
Instead, we compute polynomial fits for the scalar charge, which, after having  been derived, can be evaluated quickly and with little computational overhead.

We first construct solutions for various choices of EOS, NS mass, and scalar-tensor coupling $\alpha_0$.
We interpolate tabulated EOS data for the sly~\cite{Douchin:2001sv}, eng~\cite{Engvik:1995gn}, and H4~\cite{Lackey:2005tk} EOSs; sly is a soft EOS (compact stars) whereas H4 is relatively stiff (diffuse stars).
Then, we numerically construct NSs with masses ranging between ${m_i\in[0.5\,M_\odot,2.0\,M_\odot]}$ and scalar coupling ${\alpha_0\in[0.001,1.0]}$ and compute their scalar charge.

We calculate two types of polynomial fits of the scalar charge for each EOS.
For the first, we factor out the dominant linear dependence of $\alpha_i$ on $\alpha_0$, fitting their quotient as a fourth-order polynomial in $m_i$.
We compute the polynomial fits with least-squares regression; the fits are given in Table~\ref{FTA:tab:1dfit} for each EOS that we consider.
These fits match all of our data sets within 5\% relative error, with the greatest discrepancy arising for masses close to $2\,M_\odot$.

Although these (effectively) one-dimensional polynomial fits are crucial for some of our analysis, it is possible to construct two-dimensional fits that are simpler (fewer terms) and more accurate using more sophisticated methods.
We compute these fits using the greedy-multivariate-rational regression method developed in Ref.~\cite{London:2018nxs}.
This method relies on a greedy algorithm to construct a multivariate fit: during each iteration, it adds a polynomial term to the current fit (up to a pre-specified maximum degree) so as to best improve the agreement with the inputted data.
This process is repeated until sufficient accuracy is achieved, and then terms are systematically removed from the polynomial until the accuracy goal is saturated.
Using this method, we construct fits that agree to within 1\% relative error for each EOS---these are listed in Table~\ref{FTA:tab:2dfit}.

\subsection{Constraining $\alpha_0$ with GW170817}\label{FTA:sec:BDconstraints}

Next, we use the tools introduced previously to place constraints on the scalar-tensor coupling $\alpha_0$ in JFBD gravity with GW170817---the first GW event from a coalescing BNS.
We present two complementary analyses based off of the \FTI{} infrastructure to achieve this result.
These two methods follow the same overall approach, but adopt different statistical assumptions, utilize different waveform models, and use different numerical fits for the NS scalar charge $\alpha_i$.
In both approaches, we employ a generalized waveform model that allows for an additional contribution to the phase evolution at -1PN order, so as to reproduce the behavior seen in Eq.~\eqref{FTA:eq:DipoleTerm}; however, the parameterization of this -1PN deviation from GR differs in each approach.
Ultimately, both analyses provide a bound on $\alpha_0$ of the same order of magnitude.

The first approach we adopt directly uses the theory-agnostic constraints on a -1PN deviation, and it was originally obtained in Ref.~\cite{Abbott:2018lct}.
We use in this analysis a generalization of the tidal, aligned-spin $\texttt{SEOBNRT}$~\cite{Bohe:2016gbl,Hinderer:2016eia,Lackey:2018zvw} waveform model in which the -1PN term is parameterized by the deviation parameter $\delta \hat{\varphi}_{-2}$.
By assuming a particular NS EOS, we can use the polynomial fit in Table~\ref{FTA:tab:1dfit} in conjunction with Eq.~\eqref{FTA:eq:DipoleTerm} to map a measured value of $\delta \hat{\varphi}_{-2}$ to an inferred value on $\alpha_0$; schematically, this mapping takes the form $\alpha_0 ( \delta \hat{\varphi}_{-2}, m_1, m_2; \text{EOS})$.
Note that this mapping is infeasible using the multivariate fit in Table~\ref{FTA:tab:2dfit} because of the nonlinear dependence of $\alpha_i$ on $\alpha_0$.
Though the exact NS EOS remains unknown, we can repeat this analysis for the three candidate EOSs detailed earlier, and then use the variance on the bounds of $\alpha_0$, recovered each time, as an estimate of the systematic error arising from our ignorance of the true NS EOS.

In practice, one does not measure the masses and deviation parameter $\delta \hat{\varphi}_{-2}$ with perfect accuracy, but instead uses Bayesian inference (Sec.~\ref{sec:BayesThm}) to reconstruct the posterior distribution $P(\bm{\theta} | d, \mathcal{H})$ on these parameters given some assumed prior distribution $p(\bm{\theta}|\mathcal{H})$. So, rather than map a single point from one parameterization to another, we instead map the appropriate distributions to their counterparts in the new parameterization.
These prior and posterior distributions transform respectively as

\begin{equation}
\begin{split}
  p(\alpha_0,m_1, m_2|\mathcal{H}) = \left| \frac{\partial (\alpha_0, m_1, m_2)}{\partial (\delta \hat{\varphi}_{-2},m_1, m_2)} \right|^{-1} \\
  \times p(\delta \hat{\varphi}_{-2},m_1, m_2|\mathcal{H}), \label{FTA:eq:PriorMap}
\end{split}
\end{equation}
and
\begin{equation}
\begin{split}
  P(\alpha_0,m_1, m_2|d,\mathcal{H}) = \left| \frac{\partial (\alpha_0, m_1, m_2)}{\partial (\delta \hat{\varphi}_{-2},m_1, m_2)} \right|^{-1} \\ 
  \times P(\delta \hat{\varphi}_{-2},m_1, m_2|d,\mathcal{H}),\label{FTA:eq:PosteriorMap}
\end{split}
\end{equation}
where the first term on the right-hand side of either equations is the inverse of the Jacobian of the aforementioned transformation.

\begin{figure}
\begin{center}
\includegraphics[width=0.45\textwidth]{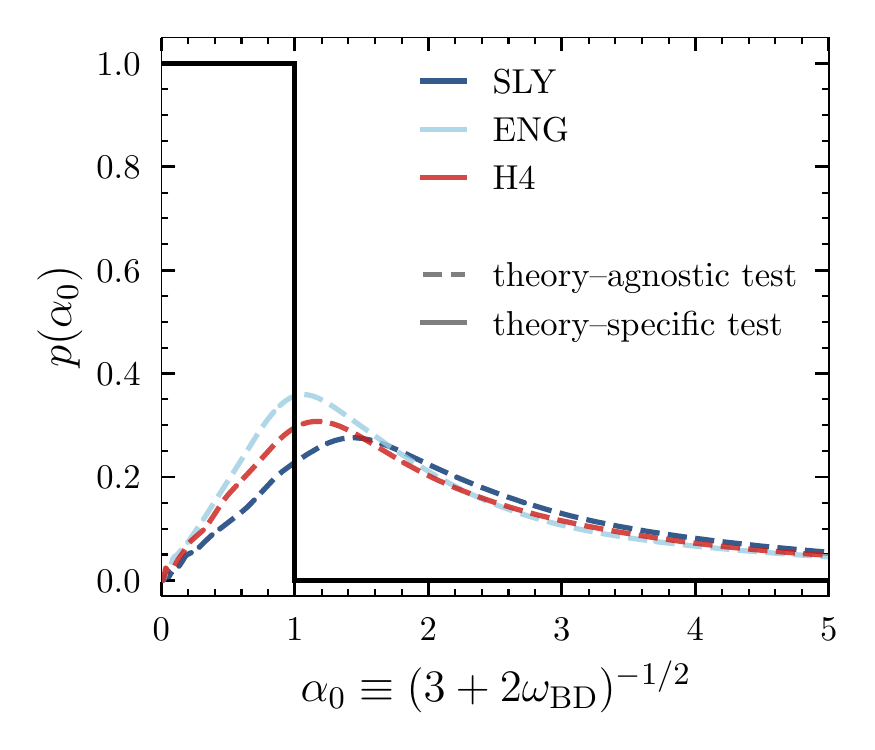}
\end{center}
\caption{Marginalized prior distributions on the JFBD parameter $\alpha_0$ used in the two analyses. The dashed colored curves depict the prior distribution equivalent to the flat prior distribution  on component masses and deviation parameter $\delta \hat{\varphi}_{-2}$ assumed in the theory-agnostic analysis. The solid black curve depicts the flat prior on $\alpha_0$ assumed in the theory-specific test.}
\label{FTA:fig:BransDickePriors} 
\end{figure}

\begin{figure}
	\begin{center}
		\includegraphics[width=0.45\textwidth]{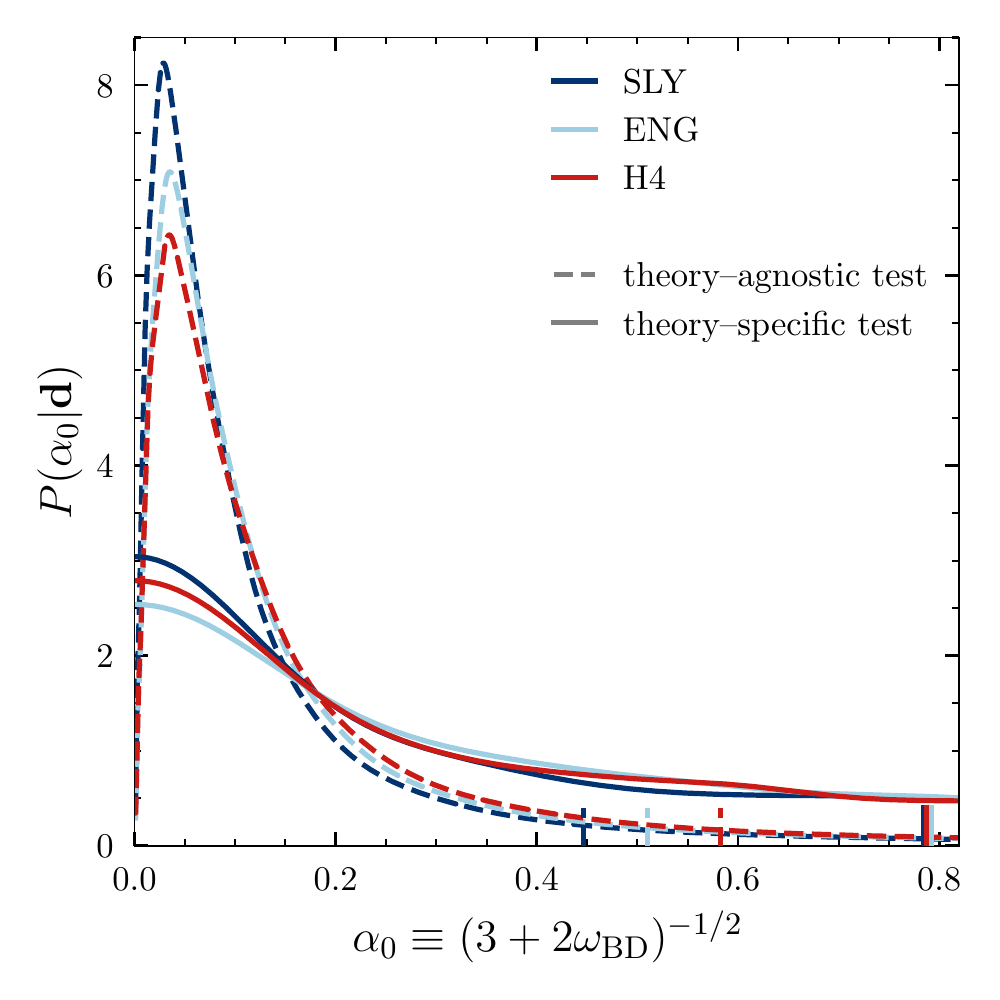}
	\end{center}
	\caption{Marginalized posterior distributions on the JFBD parameter $\alpha_0$ recovered from GW170817. The dashed curves show the posterior recovered directly from the theory-agnostic analysis. The solid curves depict the posterior using the theory-specific test of JFBD assuming a flat prior for $\alpha_0$. For both analyses, different colors correspond to different assumed EOSs. The short vertical lines on the horizontal axes represent the 90\% upper bounds on $\alpha_0$ for each analysis. The theory-agnostic test provides better bounds on  $\alpha_0$. \label{FTA:fig:BransDickePosteriors} }
\end{figure}

In the analysis of Ref.~\cite{Abbott:2018lct}, a flat prior was assumed on the component masses and deviation parameter $\delta \hat{\varphi}_{-2}$.
(We note that the upper prior bound is dictated by the theory since $\delta \hat{\varphi}_{-2} \leq 0$, see Eq.~\eqref{FTA:eq:DipoleTerm}.)
These choices reflect the theory-agnostic nature of that test; without a preferred alternative, this choice represents the simplest prior in terms of these binary parameters. Figure~\ref{FTA:fig:BransDickePriors} depicts with dashed lines how this choice of prior distribution maps to an assumed prior on $\alpha_0$ through Eq.~\eqref{FTA:eq:PriorMap}; here the different colors correspond to different assumed EOSs.
Similarly, the dashed lines in Fig.~\ref{FTA:fig:BransDickePosteriors} show the corresponding marginalized posteriors on $\alpha_0$, transformed from the posterior on $\delta \hat{\varphi}_{-2}$ and component masses via Eq.~\eqref{FTA:eq:PosteriorMap}.
This analysis provides a bound of $\alpha_0 \lesssim 2\times 10^{-1}$ ($ 5\times 10^{-1}$) at a 68\% (90\%) credible level. We find that that the systematic error arising from our ignorance of the NS EOS does not impact our estimate at 68\%, but change the bound at 90\% CL by  $\sim 30\%$. The bound on $\alpha_0$ with GW170817 is about three orders of magnitude larger than what 
obtained with the double-pulsar J0737-3039~\cite{Kramer:2021jcw}, 
which since 2003 has been tracked for about 60\,000 orbital cycles. Future observations with GW detectors on the ground and in space, will allow to reach similar or better accuracies~\cite{Perkins:2020tra}. Due to the much smaller velocity ($v/c \sim 2 \times 10^{-3}$), the double-pulsar observation becomes quickly less constraining for high PN terms entering the GW phasing (i.e., the ones discussed in Sec.~\ref{sec:results}).

{We can straightforwardly translate our bounds on the coupling to the JFBD parameter $\omega_\text{BD}$ using that $\alpha_0^2 = 1 / (3+ 2 \omega_\text{BD})$, see Table~\ref{tab:PPN}.
We find that the conservative bound in the theory-specific approach is $\omega_\text{BD} \gtrsim 1.12$ $(-0.70)$ at a 68\% (90\%) credible level. 
We also find it useful to quote here a bound on the parameterized post-Newtonian (PPN) parameter $\gamma_\text{PPN}$~\cite{Anderson:2019eay} that would correspond to our bound on $\alpha_0$, using $|1-\gamma_\text{PPN}| = 2 \alpha_0^2 / (1+\alpha_0^2)$.
This leads us to the bound $|1-\gamma_\text{PPN}| \lesssim 0.32$ $(0.77)$ at a 68\% (90\%) credible level.
However, note that our constraint originates from the dipole radiation and not from the 1PN deformation of the metric that defines $\gamma_\text{PPN}$.}

\begin{table*}\caption{{Our bounds on the coupling $\alpha_0$ translated to bounds on the JFBD parameter $\omega_\text{BD}$ and the PPN parameter $\gamma_\text{PPN}$. The numbers here indicate 68\% credible level, while the numbers in brackets are the 90\% credible level.}}\label{tab:PPN}
\begin{center}
\begin{tabular}{c@{~}r@{~~~}c@{~~~}c@{}}
  \multicolumn{4}{c}{theory-specific bounds} \\ [1ex]
  \multicolumn{2}{c}{EOS} & $\omega_\text{BD}$ & $|1-\gamma_\text{PPN}|$ \\
  \hline \\ [-2.5ex]
  sly & \cite{Douchin:2001sv} & 1.69 (-0.68) & 0.27 (0.76) \\
  eng & \cite{Engvik:1995gn}  & 1.12 (-0.70) & 0.32 (0.77) \\
  H4  & \cite{Lackey:2005tk}  & 1.20 (-0.69) & 0.31 (0.76)
\end{tabular}
\hspace{2cm}
\begin{tabular}{c@{~}r@{~~~}c@{~~~}c@{}}
  \multicolumn{4}{c}{theory-agnostic bounds} \\ [1ex]
  \multicolumn{2}{c}{EOS} & $\omega_\text{BD}$ & $|1-\gamma_\text{PPN}|$ \\
  \hline \\ [-2.5ex]
  sly & \cite{Douchin:2001sv} & 25.17 (1.00)  & 0.04 (0.33) \\
  eng & \cite{Engvik:1995gn}  & 18.69 (0.42)  & 0.05 (0.41) \\
  H4  & \cite{Lackey:2005tk}  & 14.37 (-0.02) & 0.06 (0.50)
\end{tabular}
\end{center}
\end{table*}

The second approach we employ to constrain $\alpha_0$ relies instead on a waveform model design specifically to test JFBD theory.
Using the \FTI{} infrastructure, we construct a generalized waveform model from $\texttt{SEOBNRT}$ in which the deviation parameter is precisely $\alpha_0$.
The appropriate form of the -1PN correction to the phase evolution is obtained by inserting the polynomial fit for $\alpha_i(\alpha_0,m_i)$ found in Table~\ref{FTA:tab:2dfit} for a particular choice of the EOS into Eq.~\eqref{FTA:eq:DipoleTerm}.
Additionally, unlike the previous theory-agnostic analysis in which the tidal parameters were allowed to vary freely, for this analysis, we express these parameters as functions of the respective NS masses and assumed EOS.\footnote{We use polynomial fits to the tidal parameters as a function of NS mass that are constructed in GR, not in JFBD gravity. However, the differences between these two relations scale as $\alpha_0^2$, and thus can be neglected in favor of the simpler GR relation by the same reasoning that the other sub-dominant PN effects (e.g., at 0PN order, 0.5PN order, etc.) can be ignored.}
This step reduces the dimensionality of the waveform model by two parameters while ensuring that all matter effects are handled self-consistently.
We assume a flat prior on $\alpha_0 \in [0,1]$ for this analysis; beyond this upper bound, our assumption that JFBD effects of order $\alpha_0^2$ are subdominant to the PN effects in GR is no longer valid.
This prior distribution is depicted in Fig.~\ref{FTA:fig:BransDickePriors} with a solid black curve.
Using the generalized waveform described above, we perform parameter estimation to construct the marginalized posterior distribution on $\alpha_0$, shown in Fig.~\ref{FTA:fig:BransDickePosteriors} with solid colored curves corresponding to the assumed EOS.
We obtain the upper bound of $\alpha_0 \lesssim 4\times 10^{-1}$ ($8\times 10^{-1}$) at a 68\% (90\%) credible interval where, as before, the systematic error due to ignorance of the true NS EOS does not contribute at this level of precision.
{As explained above, our constraint on $\alpha_0$ can be translated to the bounds $\omega_\text{BD} \gtrsim 14.37$ $(-0.02)$ or $|1-\gamma_\text{PPN}| \lesssim 0.06$ $(0.50)$ at a 68\% (90\%) credible level.}

Comparing the bounds set by the two analyses, we see that the theory-agnostic test provides a stronger bound on $\alpha_0$.
At first glance, this result may appear counterintuitive, as Fig.~\ref{FTA:fig:BransDickePriors} shows that this test assumed a marginalized prior on $\alpha_0$ with greater support away from zero.
The predominant cause for this discrepancy stems from how the tidal parameters are handled by each waveform model.
For the theory-agnostic test, these parameters are allowed to vary freely, independent of the masses of the NSs.
However, in the theory-specific test, the tidal parameters are linked directly to the component masses.
This latter restriction significantly affects the recovered posterior distribution on the component masses, placing much greater weight near equal-mass configurations than in the previous case.
As can be seen in Eq.~\eqref{FTA:eq:DipoleTerm}, in very symmetric configurations, the total deviation from the baseline GR waveform remains small even when $\alpha_0$ is relatively large; as a result, the JFBD parameter is more poorly measured when the tidal parameters cannot vary freely, and thus we recover a weaker bound with this theory-specific test.

\section{Conclusion}
\label{sec:conclusion}

In this paper we have developed a framework that allows us to introduce 
deviations from GR to any frequency-domain inspiral phasing, assuming 
that such corrections are small modifications to the GR signal.
For theory-agnostic tests, our \FTI{} framework has already been successfully applied to GWs 
observed by LIGO and Virgo detectors by the LVC in Refs.~\cite{TheLIGOScientific:2016src,TheLIGOScientific:2016pea,Abbott:2018lct,LIGOScientific:2019fpa,LIGOScientific:2020tif,LIGOScientific:2021sio}, while for theory-specific tests 
the \FTI{} method was employed in Ref.~\cite{Sennett:2019bpc} to set bounds 
on an effective-field theory of GR using BBH signals.

{More specifically, building on the PN SPA phasing in frequency domain, 
we have described how to apply the \FTI{} framework to multipolar 
aligned-spin waveforms, by introducing deviations parameters to GR PN terms up to 3.5PN order, and also to PN 
terms that are absent in GR, such as the -1PN and 0.5PN corrections. 
Although we apply our framework mainly to the inspiral-merger-ringdowm  \SEOBNRHM{} waveform model, 
which contains $(\ell,m) =  (2,1), (3,3), (4,4), (5,5)$ modes in addition to the dominant
  $(2,2)$ mode, the method is general and can be used for any frequency-domain waveform (or the 
Fourier-transform of a time-domain waveform). The corrections introduced in the phasing 
  are tapered off at a certain orbital frequency, which is a free
  parameter in this framework. The tapering process is introduced to
  ensure that the modified PN phase for each mode reduces to its
  corresponding GR phase during the late-inspiral--merger-ringdown stages, up to a constant 
phase shift above the tapering frequency.

{We have then discussed the application of the \FTI{} framework to BBHs, notably to  
two specific high--mass-ratio events observed by LIGO and Virgo, GW190412 and GW190814. 
The latter were previously analyzed in Ref.~\cite{LIGOScientific:2020tif}, where it was 
observed that the posterior distributions of the -1PN deviation parameter, $\delta \hat{\varphi}_{-2}$, 
were peaking away from the GR prediction, suggesting possible modeling systematic biases. Here,
  by creating simulated signals with parameters corresponding to the
  median values of the two events, we have demonstrated that these 
  features are most likely due to (unexplored) artifacts of the noise around the events rather 
than missing physics in the waveform models. Furthermore, using the simulated signals, we have showed that modes beyond the 
quadrupole \emph{do} affect the accuracy of the deviation-parameter measurements \emph{and} the GR
  parameters, especially if the signals have relatively high-mass
  ratios and inclinations. In such cases, neglecting the high modes can significantly bias the measurements at 
high SNRs, leading, erroneously, to interpret the measurement results as a
  violation of GR.}

{We have also performed robustness tests of the \FTI{}
  results, and showed that the bounds could be sensitive 
  to the tapering frequency, depending on the parameters of the signal
  being analyzed. For very low total-mass systems like the BNS 
  GW170817, the tapering frequency lies outside the frequency band where 
the majority of the SNR is accumulated, thus the bounds are not 
expected to change. On the other hand, for high--total-mass BBH systems, 
like GW190814 and GW190412, the  bounds on the deviation parameters can change. More specifically, 
bounds on deviation parameters that are measured with an accuracy of few tens of percent or less, 
may change by a factor $2\mbox{--}5$, depending on the binary's parameters, when the tapering 
frequency spans the last $3\mbox{--}4$ GW cycles before the (2,2)-mode's peak. These results suggest to 
go beyond the choice of the tapering frequency adopted so far in Refs.~\cite{TheLIGOScientific:2016src,TheLIGOScientific:2016pea,Abbott:2018lct,
LIGOScientific:2019fpa,LIGOScientific:2020tif}, where it was fixed 
to a specific value to compare with the complimentary analysis provided by the \TIGER{} framework~\cite{Agathos:2013upa}. 
In order to optimize the GR test with the \FTI{} and exploit the full SNR accumulated during the inspiral, up 
to merger, comprehensive studies would need to be undertaken to determine the best choice of the 
tapering frequency as a function of the binary's properties and the accumulated SNR. 
By contrast, we have shown that the bounds are marginally affected by the use of a different but similarly accurate waveform model, e.g., 
the state-of-the-art aligned-spin (phenomenological) \pPhenomX{} model, instead of \pSEOBNR{}.}

{Moreover, we have investigated how the presence of the deviation parameters 
in the waveform model affects the measurement of the GR parameters. We have found 
that most deviation parameters such as $\delta \hat{\varphi}_{2}$, $\delta \hat{\varphi}_{3}$, $\delta
\hat{\varphi}_{4}$, $\delta \hat{\varphi}_{5l}$, and $\delta
\hat{\varphi}_{6l}$ do not impact the GR parameters in any noticeable
way. On the other hand, the lower-order deviation parameters $\delta
\hat{\varphi}_{-2}$, $\delta \hat{\varphi}_{0}$ and $\delta
\hat{\varphi}_{1}$ affect the width of the chirp-mass measurement quite 
significantly. This is due to the fact that those 
deviation parameters are degenerate with the chirp mass. We have found that 
the extent of the correlation, however, also depends on the choice of
the tapering frequency (i.e., on the amount of SNR). The remaining 
two deviation parameters, $\delta \hat{\varphi}_{6}$ and $\delta \hat{\varphi}_{7}$, 
can cause bimodalities in the posterior distributions of the GR parameters. 
The strength of these bimodalities depends on the tapering frequency and the 
binary's parameters. We find that the bimodalities can be also caused by the fact 
that the 3PN and 3.5PN terms in the phasing are functions of the mass ratio and the spins, 
and depending on the binary's parameters, those PN coefficients can go to zero. 
It might be possible to avoid bimodalities in the GR posterior distributions by splitting the 
deviation parameters at 3PN and 3.5PN order in non-spinning and 
spinning parts, and do inference on those parts separately.}

{Finally, we have used the \FTI{} method to perform a theory-specific test 
with the BNS event GW170817 and the JFBD
gravity theory. We have obtained constraints on the JFBD parameter
  $\alpha_0$ following two strategies and employing the tidal aligned-spin 
\SEOBNRT{} waveform model. In the first approach, we directly converted the theory-agnostic -1PN
  (i.e., $\delta \hat{\varphi}_{-2}$) posterior samples into  samples
of $\alpha_0 (\delta \hat{\varphi}_{-2}, m_1, m_2; \text{EOS})$ using
the fits in Table~\ref{FTA:tab:1dfit} for different NS EOSs.  This
approach has provided us with a bound on $\alpha_0 \lesssim 2\times
10^{-1}$ ($5\times 10^{-1}$) at the 68\% (90\%) credible interval, {see also Table~\ref{tab:PPN} for the corresponding bounds on $\omega_\text{BD}$ and $\gamma_\text{PPN}$.} In the second approach, we
have employed a waveform model that is specifically designed to test JFBD,
that is we have chosen as deviation parameter directly $\alpha_0$. 
In this case, the phase corrections are given by 
Eq.~\eqref{FTA:eq:DipoleTerm} where $\alpha_1$ and $\alpha_2$ are
determined using the fits in Table~\ref{FTA:tab:2dfit}. Additionally,
in this second approach, we have also fixed the tidal parameters since 
for a given EOS they can be determined from the component masses. 
This process reduces the dimensionality of the waveform model while 
ensuring that all matter effects are handled self-consistently. Using a flat prior on
$\alpha_0\in [0,1]$, we have found a bound on
$\alpha_0 \lesssim 4\times 10^{-1}$ ($8\times 10^{-1}$) at 
the 68\% (90\%) credible interval, {see also Table~\ref{tab:PPN} for the corresponding bounds on $\omega_\text{BD}$ and $\gamma_\text{PPN}$.}
The main reason for the difference in the bounds can be traced to how the tidal parameters are handled in
the two approaches. In the theory-agnostic approach the tidal
parameters were allowed to vary freely during the inference analysis. By contrast, 
in the theory-specific approach, tidal parameters and scalar
  sensitivities cannot be treated as independent, but must be computed
  consistently (for each NS mass) by fixing the EOS. Thus,
theory-specific and theory-agnostic analyses test slightly different
statistical hypotheses, and within a fully Bayesian framework,
converting results from one to the other requires care. To our knowledge, 
the impact of statistical hypotheses on relating theory-specific and 
theory-agnostic bounds has not been studied in detail in the context of 
GW tests of GR, and thus offers an interesting new avenue for future work.}

{Lastly, the \FTI{} framework can be applied to perform 
theory-specific tests with non-GR theories other that JFBD gravity, as done 
in Ref.~\cite{Sennett:2019bpc}. Additionally, the framework can be adapted to constrain other effects
that leave an impact on the inspiral phase --- for example the existence of exotic compact
objects having a spin-induced quadrupole moment different from the one 
of a BH in GR~\cite{Krishnendu:2017shb,Narikawa:2021pak}. However, 
more crucial for future, more sensitive observations, is the extension of the 
\FTI{} framework to the spin-precessing case. This will be  possible by 
applying the frequency-domain corrections to the phasing in the co-precessing frame, where the GWs
are usually well approximated by aligned-spin waveforms~\cite{Buonanno:2002fy,Ossokine:2020kjp}.}

\section*{Acknowledgements}

{The authors thank Michalis Agathos, Stanislav Babak, Richard Brito, Walter Del Pozzo, and Sylvain Marsat for useful discussions and/or involvement in building the \FTI{} infrastructure in the LIGO Algorithm Library and/or employing 
it with LIGO-Virgo observations. They are grateful to Laszlo Gergely, Max Isi and B. Sathyaprakash for reviewing the results presented in Sec.~\ref{sec:conclusion}, which were originally obtained by Noah Sennett as part of the LIGO-Virgo analysis of GW170817~\cite{Abbott:2018lct}. We thank Norbert Wex for providing us with the most recent bounds on $\alpha_0$ from the double pulsar. The authors also thank Marta Colleoni for carefully reading the manuscript and providing 
useful comments. The authors are grateful for the computational resources provided by the LIGO Laboratory, which is 
supported by the National Science Foundation (NSF) under Grants PHY0757058 and PHY-0823459, as well as,  
the computational resources at the Max Planck Institute for Gravitatinal Physics in Potsdam, specifically the 
high-performace computing cluster {\tt{Hypatia}}. The material presented in this manuscript is based upon work supported by NSF’s LIGO Laboratory, which is a major facility fully funded by the NSF. Finally, the authors would like to thank everyone at the frontline of the Covid-19 pandemic.} 

{This research has made use of data, software and/or web tools obtained from the Gravitational Wave Open Science Center (https://www.gw-openscience.org), a service of LIGO Laboratory, the LIGO Scientific Collaboration and the Virgo Collaboration, and Zenodo (https://zenodo.org/record/5172704)}.


\appendix

\section{The 3.5 PN phasing in the stationary-phase approximation}
\label{appendix:3.5PNPhasing}

Here, we write the PN coeffcients entering the GW phasing in the SPA approximation~\cite{Sathyaprakash:1991mt} through 3.5PN order, including spin effects in GR. 

Let us introduce the following quantities:
\begin{subequations}
\begin{align}
&\delta = \dfrac{(m_1 - m_2)}{M}= 1-4\nu, \\
& \chi_{\rm S} = \dfrac{(\chi_1 +\chi_2)}{2},\\
& \chi_{\rm A} = \dfrac{(\chi_1 -\chi_2)}{2}, 
\end{align}
\end{subequations}
and the Euler's constant $\gamma_{\rm E}$. The PN coefficients in Eq.~\eqref{FTA:eq:inspPhase} read~\cite{Buonanno:2009zt, Arun:2008kb, Bohe:2013cla}:
\begin{widetext}
\begin{subequations}
	\begin{align}
	& \psi_0 = 1, \\ 
	& \psi_1 = 0, \\
	& \psi_2 =  \dfrac{3715}{756}  + \dfrac{55 \nu}{9},\\
	& \psi_3 = -16 \pi +\frac{113 \delta\,\chi_{\rm A}}{3}+\left(\frac{113}{3}-\frac{76 \nu }{3}\right) \chi_{\rm S}, \\
	& \psi_4 = \frac{15293365}{508032}+\frac{27145 \nu }{504}+\frac{3085 \nu ^2}{72}+\left(-\frac{405}{8}+200 \nu \right) \chi_{\rm A}^2-\frac{405 \delta\,\chi_{\rm A}\,\chi_{\rm S}}{4}+\left(-\frac{405}{8}+\frac{5 \nu }{2}\right) \chi_{\rm S}^2,\\
	& \psi_5 =\frac{38645 \pi }{756}-\frac{65 \pi  \nu }{9}+ \left(-\frac{732985}{2268}-\frac{140 \nu }{9}\right)\delta\, \chi_{\rm A}+\left(-\frac{732985}{2268}+\frac{24260 \nu
	}{81}+\frac{340 \nu ^2}{9}\right) \chi_{\rm S},\\
	&\psi_{5l} = 3\psi_5 = \frac{38645 \pi }{252}-\frac{65 \pi  \nu }{3}+ \left(-\frac{732985}{756}-\frac{140 \nu }{3}\right)\delta\,\chi_{\rm A}+\left(-\frac{732985}{756}+\frac{24260 \nu
	}{27}+\frac{340 \nu ^2}{3}\right) \chi_{\rm S}, \\
	&\psi_6 = \frac{11583231236531}{4694215680}-\frac{6848 \log(4)}{21}-\frac{640 \pi ^2}{3}+\frac{6848 \gamma_{\rm E} }{21}+\left(-\frac{15737765635}{3048192}+\frac{2255 \pi
		^2}{12}\right) \nu +\frac{76055 \nu ^2}{1728}-\frac{127825 \nu ^3}{1296} \nonumber\\
	& \hspace{0.8cm}+\frac{2270 \pi  \delta\,\chi_{\rm A}}{3} + \left(\frac{2270 \pi }{3}-520 \pi  \nu \right )\chi_{\rm S} +\left(\frac{75515}{288}-\frac{547945 \nu
	}{504}-\frac{8455 \nu ^2}{24}\right) \chi_{\rm A}^2+ \left (\frac{75515}{144}-\frac{8225 \nu }{18}\right) \delta\,\chi_{\rm A}\,\chi_{\rm S} \nonumber\\
	&\hspace{0.8cm}+\left(\frac{75515}{288}-\frac{126935 \nu }{252}+\frac{19235 \nu ^2}{72}\right) \chi_{\rm S}^2, \label{psi6exp} \\
	&\psi_{6l}= -\frac{6848}{21}, \label{psi6lexp} \\
	&\psi_7 = \frac{77096675 \pi }{254016}+\frac{378515 \pi  \nu }{1512}-\frac{74045 \pi  \nu ^2}{756}+ \left(-\frac{25150083775}{3048192}+\frac{26804935 \nu }{6048}-\frac{1985
		\nu ^2}{48}\right)\delta\, \chi_{\rm A} \nonumber\\
	&\hspace{0.8cm}+\left(-\frac{25150083775}{3048192}+\frac{10566655595 \nu }{762048}-\frac{1042165 \nu ^2}{3024}+\frac{5345 \nu ^3}{36}\right) \chi_{\rm S} . \label{psi7exp}
	\end{align}	
\end{subequations}
\end{widetext}

\bibliography{FTA}
\end{document}